\documentclass[twocolumn]{aastex701}

\usepackage{amsmath}
\shorttitle{ALMA View of GoHam}
\shortauthors{Law et al.}

\begin{document}

\title{The ALMA View of the Edge-on Gomez's Hamburger System: A Highly-Dynamic, Asymmetric Protoplanetary Disk Reveals the Earliest Phases of Giant Planet Formation}

\correspondingauthor{Charles J.\ Law}
\email{law06855@umn.edu}

\author[0000-0003-1413-1776]{Charles J.\ Law}
\affiliation{Minnesota Institute for Astrophysics, University of Minnesota, 116 Church St. SE, Minneapolis, MN 55455}
\affiliation{Department of Astronomy, University of Virginia, Charlottesville, VA 22904, USA}
\email{cjl8rd@virginia.edu}

\author[0009-0007-5371-3548]{Jensen Lawrence}
\affiliation{Department of Earth, Atmospheric, and Planetary Sciences, Massachusetts Institute of Technology, Cambridge, MA 02139, USA}
\email{jptl@mit.edu}

\author[0000-0003-1534-5186]{Richard Teague}
\affiliation{Department of Earth, Atmospheric, and Planetary Sciences, Massachusetts Institute of Technology, Cambridge, MA 02139, USA}
\email{rteague@mit.edu}

\author[0000-0001-7233-4171]{Zhe-Yu Daniel Lin}
\affiliation{Department of Astronomy, University of Virginia, Charlottesville, VA 22904, USA}
\email{zlin@carnegiescience.edu}

\author[0000-0003-1837-3772]{Romane Le Gal}
\affiliation{Université Grenoble Alpes, CNRS, IPAG, F-38000 Grenoble, France}
\affiliation{Institut de Radioastronomie Millimetrique (IRAM), 300 rue de la piscine, F-38406 Saint-Martin d’Hères, France}
\email{romane.le-gal@univ-grenoble-alpes.fr}

\author[0000-0002-5688-6790]{Kristina Monsch}
\affiliation{Center for Astrophysics \textbar\, Harvard \& Smithsonian, 60 Garden St., Cambridge, MA 02138, USA}
\email{kristina.monsch@cfa.harvard.edu}

\author[0000-0003-3122-4894]{Gordian Edenhofer}
\affiliation{Max Planck Institute for Astrophysics, Karl-Schwarzschild-Strasse 1, 85748, Garching bei München, Germany}
\email{gordian.edenhofer@gmail.com}

\author[0000-0002-0880-8296]{Hans Baehr}
\affiliation{Max-Planck-Institut für Astronomie (MPIA), Königstuhl 17, 69117 Heidelberg, Germany}
\email{baehr@mpia.de}

\author[0000-0003-4179-6394]{Edwin A. Bergin}
\affiliation{Department of Astronomy, University of Michigan, 1085 S. University Ave, Ann Arbor, MI 48109, US}
\email{ebergin@umich.edu}

\author[0000-0003-2076-8001]{L. Ilsedore Cleeves}
\affiliation{Department of Astronomy, University of Virginia, Charlottesville, VA 22904, USA}
\email{lic3f@virginia.edu}

\author[0000-0003-2341-5922]{Anne Dutrey}
\affiliation{Univ. Bordeaux, CNRS, Laboratoire d’Astrophysique de Bordeaux, UMR 5804, F-33600 Pessac, France}
\email{anne.dutrey@u-bordeaux.fr}

\author[0009-0003-8967-9522]{Coralie Foucher}
\affiliation{Laboratoire d’Astrophysique de Bordeaux, Université de Bordeaux, CNRS, B18N, Allée Geoffroy Saint-Hilaire, F-33615 Pessac}
\email{coralie.foucher@u-bordeaux.fr}

\author[0000-0003-3773-1870]{Stéphane Guilloteau}
\affiliation{Univ. Bordeaux, CNRS, Laboratoire d’Astrophysique de Bordeaux, UMR 5804, F-33600 Pessac, France}
\email{stephane.guilloteau@u-bordeaux.fr}

\author[0000-0002-9593-7618]{Thomas J. Haworth}
\affiliation{Astronomy Unit, School of Physics and Astronomy, Queen Mary University of London, London E1 4NS, UK}
\email{t.haworth@qmul.ac.uk}

\author[0000-0002-1493-300X]{Thomas Henning}
\affiliation{Max-Planck-Institut für Astronomie (MPIA), Königstuhl 17, 69117 Heidelberg, Germany}
\email{henning@mpia.de}

\author[0000-0003-1008-1142]{John D. Ilee}
\affiliation{School of Physics and Astronomy, University of Leeds, Leeds, UK, LS2 9JT}
\email{J.D.Ilee@leeds.ac.uk}

\author[0000-0001-6684-6269]{Marija R. Jankovic}
\affiliation{Institute of Physics Belgrade, University of Belgrade, Pregrevica 118, 11080 Belgrade, Serbia}
\email{marija.jankovic@ipb.ac.rs}

\author[0000-0002-6429-9457]{Kamber R. Schwarz}
\affiliation{Max-Planck-Institut für Astronomie (MPIA), Königstuhl 17, 69117 Heidelberg, Germany}
\email{schwarz@mpia.de}

\author[0000-0001-7962-1683]{Ilaria Pascucci}
\affiliation{Lunar and Planetary Laboratory, The University of Arizona, Tucson, AZ 85721, USA}
\email{pascucci@arizona.edu}

\author[0000-0002-3913-7114]{Dmitry Semenov}
\affiliation{Max-Planck-Institut für Astronomie (MPIA), Königstuhl 17, 69117 Heidelberg, Germany}
\affiliation{Zentrum für Astronomie der Universität Heidelberg, Institut für Theoretische Astrophysik, Albert-Ueberle-Str. 2, 69120 Heidelberg, Germany}
\email{semenov@mpia.de}

\author[0000-0002-0661-7517]{Ke Zhang}
\affiliation{Department of Astronomy, University of Wisconsin-Madison, 475 N Charter St, Madison, WI 53706}
\email{ke.zhang@wisc.edu}

\begin{abstract}

Chemical tracers provide some of the strongest observational signatures of ongoing planet formation and localized dynamical perturbations in protoplanetary disks. In particular, sulfur-bearing molecules are predicted to be enhanced in regions of shock heating, ice sublimation, and gravitational instability. Here, we present high-angular-resolution ($\approx$0\farcs2) Atacama Large Millimeter/submillimeter Array observations of $^{12}$CO J=3--2, $^{13}$CO J=3--2, CS J=7--6, and SO J$_{\rm N}$=8$_8$--7$_7$ toward the large, edge-on Gomez's Hamburger~(`GoHam'; IRAS~18059-3211) disk. We detect a narrow, one-sided arc of SO emission that peaks near a previously-identified gas over-density, suggesting localized heating around an early-stage giant protoplanet or disk fragment. The edge-on geometry of GoHam enables us to place this chemical signature in the broader context of the disk gas and dust structure. To do so, we map the vertical distribution of molecular gas relative to millimeter- and (sub)-micron-sized dust, identify a pronounced north-south continuum asymmetry, and detect non-Keplerian $^{12}$CO and $^{13}$CO emission indicative of a disk wind. We also derive a dynamical stellar mass of 2.2~$\pm$~0.5 M$_{\odot}$ and a revised dust-extinction-map-based distance of 139~$\pm$~24~pc, which places GoHam in the outskirts of the Scorpius-Centaurus association. Together, these observations reveal a highly dynamic disk in which localized sulfur chemistry may trace one of the earliest observable stages of wide-separation giant planet formation.
\end{abstract}



\section{Introduction} \label{sec:intro}

One of the major challenges in planet formation is identifying observational signatures of embedded protoplanets during their earliest stages of growth. Chemical tracers provide a powerful avenue for doing so, because nascent planets, disk fragments, and associated shocks can locally alter the thermal structure of the disk, driving ice sublimation and rapid changes in gas-phase abundances \citep[e.g.,][]{Cleeves15,Ilee17}. As a result, localized chemical enhancements may reveal ongoing planet formation even when the underlying source remains difficult to detect directly \citep[e.g.,][]{Andrews21, Cugno25}. 

Recent Atacama Large Millimeter/submillimeter Array (ALMA) observations have demonstrated the potential of sulfur-bearing molecules as probes of dynamical activity in protoplanetary disks, including infalling material from the surrounding environment \citep{Tang14, Garufi22, Huang23}, spiral density waves \citep{Zagaria25, Temmink26}, gravitational instabilities \citep{Speedie25}, and perturbations associated with embedded companions \citep{Booth23_HD100546, Law23_HD16, Yoshida24, Yoshida26}. Species such as SO, SO$_2$, H$_2$S, OCS, and SiS may be enhanced by these processes, which liberate sulfur-bearing material from ices and drive subsequent gas-phase chemistry \citep{Pineau93, Podio15, Holdship19, vanGelder21, Fortenberry24}. Among these molecules, SO is especially promising, because its abundance responds to both thermal and shock processing, and it has now been detected in a growing number of Class~II disks \citep[e.g.,][]{Pacheco16, Dutrey24, Zagaria25, Booth26, Yamato26}. Models of gravitationally-unstable disks also predict strong SO enhancements around forming fragments and accreting clumps \citep{Ilee11, Ilee17}, providing a potential opportunity to trace the onset of fragment collapse and subsequent protoplanet formation.

Interpreting localized chemical signatures requires knowledge of the surrounding disk, particularly its vertical density, temperature, irradiation, and kinematic structure. These properties determine the chemical and excitation conditions throughout the disk and, for highly-inclined systems, strongly affect how emission from different disk heights is projected along the line of sight. High-resolution ALMA observations have revealed a wide range of disk surface perturbations, including substructures, non-axisymmetric morphologies, kinematic deviations, and meridional flows that are frequently interpreted as evidence for embedded planets \citep[e.g.,][]{Pinte18_planet, Pinte19, Teague19Natur, Law21, Paneque21, Galloway23, Pinte25}. However, without constraints on the local three-dimensional structure, it can be difficult to distinguish genuine chemical enhancements from variations in the underlying physical conditions or kinematics. While new techniques have enabled measurements of gas emission surfaces in an increasing number of moderately inclined (${\approx}$30-75$^\circ$) disks \citep[e.g.,][]{Rosenfeld13, pinte18, disksurf_Teague, Law21, Law22, Izquierdo22, Stapper23, Paneque23, Law23, Law24, Kurtovic24, Galloway25}, these systems still suffer from projection effects, potentially complicating the interpretation of vertically-structured emission.

Instead, disks viewed at, or near, edge-on orientation offer a clear geometric advantage, as their vertical gas and dust layers are spatially separated and directly observable \citep[][]{Dutrey17, Podio20, Flores21, RR21, Villenave22, Dutrey25, Guilloteau25, Foucher25}. By allowing for an unambiguous view of their vertical morphology, edge-on disks provide essential benchmarks for testing models of disk-planet interactions and interpreting a variety of chemical, dynamical, and morphological signatures within a common geometric framework. Among edge-on systems, Gomez's Hamburger is particularly well-suited for such a study, due to its bright, gas-rich, and spatially-extended disk \citep{Bujarrabal08, Teague20_goham, Cusson26}. It also has existing evidence of ongoing planet formation in the form of a localized gas overdensity in the southern half of the disk that may represent a several-Jupiter-mass fragment formed through gravitational instability \citep{Bujarrabal09, Berne15}.

In this paper, we present high-resolution ALMA observations of Gomez's Hamburger. We detected highly-asymmetric SO emission that is spatially associated with a previously-identified gas over-density, which is consistent with localized heating near a forming giant-planet or disk fragment. We leveraged the edge-on geometry of this system to place this chemical signature within the broader context of its disk structure, as traced by the 0.9~mm continuum and $^{12}$CO, $^{13}$CO, and CS emission lines, which reveal several dynamical processes shaping its gas and dust distributions. In Section \ref{sec:go_ham_source}, we describe the Gomez's Hamburger system and list the observational details in Section \ref{sec:observations_overview}. We present our results in Section \ref{sec:results}, including mapping the continuum and line emission morphology and estimating the dynamical stellar mass. In Section \ref{sec:discussion}, we discuss the origin of the observed emission signatures in GoHam, and we summarize our conclusions in Section \ref{sec:conlcusions}.

\section{The GoHam Disk}
\label{sec:go_ham_source}

The Gomez’s Hamburger (hereafter `GoHam') system comprises an A-type pre-main-sequence star IRAS~18059-3211 and an edge-on protoplanetary disk. The distance to GoHam has long been uncertain due to its edge-on nature, which obscures the central star and complicates reliable astrometric measurements. Here, we adopt a newly-revised distance estimate of 139 $\pm$ 24~pc\footnote{Throughout the text, we adopt this distance and update physical scales previously referred to in the literature when assuming a far distance of 250~pc \citep[e.g.,][]{Teague20_goham}. Notably, this significantly reduces prior physical size estimates of the GoHam disk and lowers previous distance-dependent disk-mass estimates by a factor of a few.} based on 3D interstellar extinction maps, which places the system in the outskirts of the Scorpius-Centaurus (Sco-Cen) association (see Appendix \ref{sec:app:revised_distance} for details). While GoHam was originally classified as an evolved A0-star surrounded by a planetary nebula \citep{Ruiz87}, CO isotopologue line observations with the Submillimeter Array (SMA) showed the presence of Keplerian-rotating gas \citep{Bujarrabal08, Teague20_goham}, and Hubble Space Telescope (HST)/NICMOS scattered light observations revealed the characteristic flared geometry associated with small dust grains in the upper layers of a protoplanetary disk \citep{Wood08, Bujarrabal09}. 

GoHam hosts one of the largest protoplanetary disks identified to date with a $^{12}$CO radial extent of ${\approx}$6\farcs5 (${\approx}$900~au) and vertical height up to ${\approx}$2\farcs5 (${\approx}$350~au) \citep{Teague20_goham}. Based on disk structure models, the GoHam disk is massive (up to a few tenths of M$_{\odot}$) \citep{Wood08, Bujarrabal08, Bujarrabal09} and may be marginally gravitationally unstable with an inferred Toomre parameter of Q $\lesssim$ 2 \citep{Berne15}. GoHam is also known to host a localized over-density, referred to as GoHam~b, in the southern half of the disk at a projected distance of ${\approx}$2$^{\prime \prime}$ (${\approx}$280~au) that is thought to be forming due to gravitational instability (GI). GoHam~b has been identified in the form of asymmetric excess emission in $^{13}$CO \citep{Bujarrabal09, Teague20_goham} and a local reduction in 8.6~$\mu$m and 11.2~$\mu$m PAH emission \citep{Berne15}. Because PAHs are excited by stellar UV radiation, this localized PAH deficit can arise from enhanced UV shielding by a dense gas overdensity \citep{Berne15}. Together with the excess $^{13}$CO emission, these observations are consistent with the presence of a cold, dense gas clump likely containing a mass of at least several Jupiter masses in a spherical region with a radius no larger than ${\approx}$0\farcs6 (${\approx}$80~au) \citep{Berne15, Teague20_goham}. While GoHam~b is a promising candidate for a young protoplanet formed by GI, the exact nature of this clump and a robust determination of its properties have been limited by the lack of existing high-angular resolution observations.

Given its edge-on geometry, large disk size, and the potential presence of an embedded, nascent wide-separation protoplanet, subarcsecond observations of GoHam offer us a powerful opportunity to map the gas structure of this unique planet-forming disk in detail. This paper presents a detailed description of newly-acquired ALMA data, which we use to establish an empirical overview of the detailed gas and dust morphology in GoHam. A forthcoming companion paper \citep{jensen_goham} uses these data to then perform a detailed tomographic reconstruction of the three-dimensional disk structure, including deriving maps of gas temperature and isotopic ratios.

\begin{deluxetable*}{cccccccccccc}
\tablecaption{Details of ALMA Observations\label{tab:full_obs_program_details}}
\tablewidth{0pt}
\tablehead{
\colhead{UT Date} & \colhead{No. Ants.} & \colhead{Int. Time} & \colhead{Baseline Range} & \colhead{Res.} & \colhead{MRS} & \colhead{PWV} & \multicolumn3c{Calibrators} \\ \cline{8-10} 
\colhead{} & \colhead{} & \colhead{(min)} & \colhead{(m)} & \colhead{($^{\prime \prime}$)} & \colhead{($^{\prime \prime}$)}  & \colhead{(mm)} & \colhead{Flux}  & \colhead{Bandpass} & \colhead{Phase} 
}
\startdata
2022-10-15 & 44 & 44.6 & 15.1 -- 483.9 & 0.5 & 5.7 & 0.7 & J1924-2914 & J1924-2914 & J1733-3722 \\
2024-06-05 & 44 & 50.1 & 15.1 -- 1397.8 & 0.2 & 2.5 & 1.3 & J1924-2914 & J1924-2914 & J1802–3940 \\
2024-06-11 & 43 & 50.1 & 15.1 -- 1397.8 & 0.2 & 2.5 & 0.7 & J1924-2914 & J1924-2914 & J1802–3940 \\
2024-06-20 & 45 & 50.1 & 15.1 -- 2046.8 & 0.1 & 2.0 & 1.2 & J1924-2914 & J1924-2914 & J1802–3940
\enddata
\end{deluxetable*}

\section{Observations}
\label{sec:observations_overview}

\subsection{Observational Details} \label{sec:observations_details}

We present Band 7 observations of GoHam from ALMA project 2022.1.00269.S (PI: C. Law), which included short-baseline (SB) and long-baseline (LB) data. The SB data comprised one execution block (EB) observed on 15 October 2022 for an on-source time integration of 44.6~min using 44 antennas with a projected baseline range of 15.1--483.9~m. The mean PWV was 0.7~mm and maximum recoverable scale (MRS) was 5\farcs7. The LB data comprised three EBs observed from 5 June 2024 to 20 June 2024 and used between 43 and 45 antennas for a total on-source integration time of 150.3~min. Projected baselines ranged from 15.1--2046.8~m and the MRS was ${\approx}$2\farcs0-2\farcs5 with a typical PWV of 0.7-1.3~mm. The quasar J1924-2914 was used for flux and bandpass calibration for all observations, while J1733-3722 and J1802–3940 were the phase calibrators for the SB and LB observations, respectively. For all EBs, the correlator was set up to observe $^{12}$CO J=3--2 at ${\approx}$13~m~s$^{-1}$ (31~kHz) and $^{13}$CO J=3--2, CS J=7--6, SO J$_{\rm{N}}$=8$_8$--7$_7$, and CH$_3$CN J=18--17 (K=0,1) at ${\approx}$27-28~m~s$^{-1}$ (61~kHz). Here, we focus on all but the latter CH$_3$CN lines. One broad (1.875~GHz) spectral window was dedicated to the continuum at coarse velocity resolution (${\approx}$1~km~s$^{-1}$) and was centered at 333.0~GHz (0.9~mm). Table \ref{tab:full_obs_program_details} provides detailed information about each observation. 

\subsection{Self Calibration and Imaging} \label{sec:self_cal_obs}

The data were initially calibrated by the ALMA staff using the ALMA calibration pipeline and the required version of CASA \citep{McMullin_etal_2007, CASATeam20}. Self-calibration and all subsequent imaging and analysis used CASA \texttt{v6.6.3}.

We performed self calibration closely following the procedures outlined in \citet{Oberg21_MAPSI}. Briefly, we flagged line emission in each spectral window to generate pseudo-continuum visibilities, which were then combined with the continuum-only spectral window. Each execution block was aligned to a common phase center by fitting a Gaussian profile with the \texttt{imfit} task and using the \texttt{fixvis} and \texttt{fixplanets} tasks. We first performed two rounds of phase (\texttt{solint}=`inf', 900s) and one round of amplitude (\texttt{solint}=`inf') self-calibration on the SB data, which resulted in a ${\approx}$6$\times$ improvement in the peak continuum signal-to-noise ratio (SNR). Then, we concatenated the self-calibrated SB data with the LB data, and the combined visibilities were self-calibrated together with three rounds of phase-only (\texttt{solint}=`inf', 360s, 180s) self-calibration for a ${\approx}$3$\times$ improvement in the peak SNR of the continuum. We ultimately applied the resulting calibration solutions to the unflagged visibilities, before subtracting the continuum with a first-order polynomial using the \texttt{uvcontsub} task.

\begin{figure*}[!htpb]
\centering
\includegraphics[width=\linewidth]{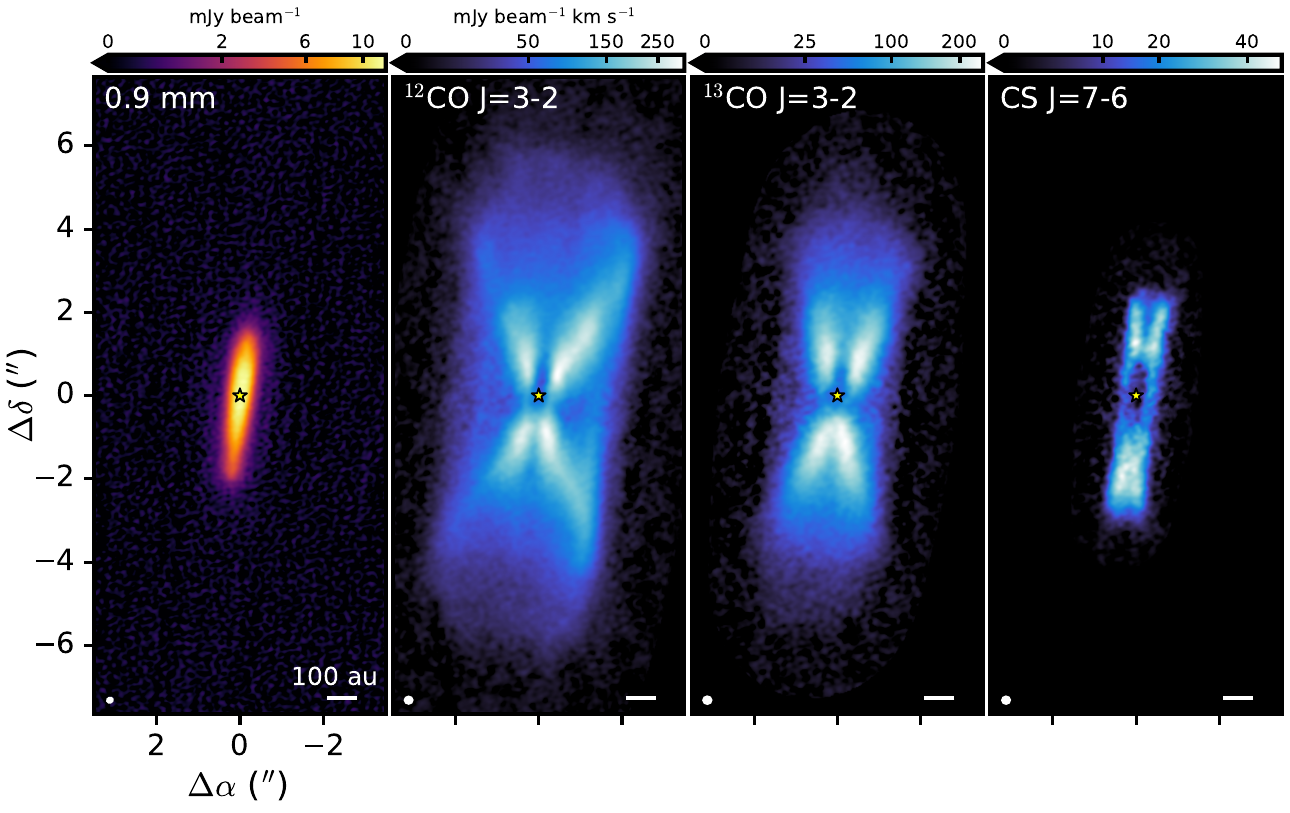}
\vspace{-20pt}
\caption{0.9~mm continuum image and zeroth moment maps of $^{12}$CO J=3--2, $^{13}$CO J=3--2, and CS J=7--6 in the GoHam disk. Color stretches have been applied to highlight faint, outer disk emission, and the central star position is marked. The synthesized beam and a scale bar indicating 100~au is shown in the lower left and right corner, respectively, of each panel.}
\label{fig:mom0_gallery}
\end{figure*}

We used the \texttt{tclean} task to produce images of the $^{12}$CO J=2--1, $^{13}$CO J=2--1, CS J=7--6, and SO J$_{\rm{N}}$=8$_8$--7$_7$ lines with Briggs weighting and a Keplerian mask generated with the \texttt{keplerian\_mask} \citep{Teague20_Keplerian_mask} code. Each mask was based on the stellar and disk parameters of GoHam and was visually inspected to ensure that it contained all emission present in the channel maps. In a few cases, the mask required minor manual adjustments by increasing, e.g., maximum radius, beam convolution size, to include low-level emission extending beyond the initial Keplerian mask. For example, the mask for $^{12}$CO was expanded to include diffuse emission toward the northern side of the disk. We adopted a Briggs \texttt{robust} parameter of 0.5 for all images to balance achieving high SNRs, while maintaining sufficient angular resolution. All images were generated at the native velocity resolution, with the exception of SO, which was binned to 100~m~s$^{-1}$ to improve its SNR. All line images were made using the ‘multi-scale’ deconvolver with pixel scales of [0,5,15,25,50,100] and were CLEANed down to a 4$\sigma$ level, where $\sigma$ was the root-mean-square (RMS) noise measured in a line-free channel of the dirty image. Depending on the line and channel spacing, typical beam size and RMS values are ${\approx}$0\farcs17-0\farcs20 and ${\approx}$4-6~mJy~beam$^{-1}$, respectively. 

We also made a 0.9~mm continuum image from the dedicated continuum window combined with the full bandwidth of the observations after flagging channels containing line emission. We used an elliptical mask during CLEANing. After experimenting with different values, we adopted \texttt{robust}$=-0.5$ for the final continuum image. This choice provided the ideal compromise between maximizing angular resolution and not resolving out the major axis (see Appendix \ref{sec:app:robust_continuum_imaging}). Otherwise, the CLEANing parameters were the same. Table \ref{tab:image_info} summarizes all image properties. The continuum RMS is 0.12~mJy~beam$^{-1}$ with a beam size of 0\farcs14$\times$0\farcs12 and position angle of $-$70.8$^{\circ}$.

We then applied the so-called ‘JvM’ correction \citep{JvM95, Czekala21} to all image cubes to scale image residuals by $\epsilon$, the ratio of the effective areas of the CLEAN and dirty beams. This ensures that the final images are in units consistent with the CLEAN model. Table \ref{tab:image_info} lists all $\epsilon$ values. While we used the JvM-corrected images here, we also verified that line emission morphologies extracted from either the JvM- or non-JvM-corrected images yield consistent results.

\setlength{\tabcolsep}{3.5pt}
\begin{deluxetable*}{lcccccccccccccc}
\tabletypesize{\footnotesize}
\tablecaption{Image Cube and Line Properties\label{tab:image_info}}
\tablehead{
\colhead{Transition} & \colhead{Freq.} & \colhead{Beam} & \colhead{JvM $\epsilon$\tablenotemark{a}} & \colhead{\texttt{robust}} & \colhead{Chan. $\delta$v} & \colhead{RMS} & \colhead{E$_{\rm{u}}$} & \colhead{A$_{\rm{ul}}$} & \colhead{g$_{\rm{u}}$} & \colhead{Int. Flux\tablenotemark{b}}  \\ 
\colhead{} & \colhead{(GHz)} & \colhead{ ($^{\prime \prime} \times ^{\prime \prime}$, $\deg$)} & & & \colhead{(m~s$^{-1}$)} & \colhead{(mJy~beam$^{-1}$)} & \colhead{(K)} & \colhead{($\log_{10}$ s$^{-1}$)} & & \colhead{(Jy~km~s$^{-1}$}) }
\startdata
0.9~mm cont.        & 333.000000 & 0.14~$\times$~0.12, $-$70.8 & 0.97 & $-$0.5 & \ldots & 0.12 & \ldots & \ldots & \ldots &   0.881~$\pm$~0.003 \\
$^{12}$CO J=3--2        & 345.795990 & 0.19~$\times$~0.17, $-$88.7 & 0.54 & 0.5 & 13.2 & 6.37  & 33 & $-$5.603 & 7 &  83.6 $\pm$ 2.3 \\
$^{13}$CO J=3--2 & 330.587965 & 0.20~$\times$~0.18, $-$88.0 & 0.53 & 0.5 & 27.7 & 5.48 & 32 & $-$5.960 & 14 & 40.6 $\pm$ 1.7  \\
CS J=7--6 & 342.882850 & 0.19~$\times$~0.18, $-$87.9 & 0.52 & 0.5 & 26.7 & 3.84 & 66 & $-$3.077 & 15 & 3.3 $\pm$ 0.28 \\
SO J$_{\rm{N}}$=8$_{8}$--7$_{7}$ & 344.310612 & 0.19~$\times$~0.18, $-$85.9 & 0.53 & 0.5 & 100.0 & 2.18 & 87 & $-$3.285 &17 & 0.20 $\pm$ 0.08 \\
\enddata
\tablecomments{The spectroscopic constants for all lines are taken from the CDMS database \citep{Muller01, Muller05, Endres16}.}
\tablenotetext{a}{The ratio of the CLEAN beam and dirty beam effective area used to scale image residuals to account for the effects of non-Gaussian beams. See Section \ref{sec:self_cal_obs} and \citet{JvM95, Czekala21} for further details.}
\tablenotetext{b}{Uncertainties are derived via bootstrapping  and do not include the systematic calibration flux uncertainty (${\sim}$10\%). Uncertainties evaluated on the non-JvM-corrected cubes, as RMS can be underestimated by the JvM correction \citep{Casassus22}. The continuum flux has units of Jy.}
\end{deluxetable*} \vspace{-14pt} \setlength{\tabcolsep}{4pt}

\subsection{Moment Maps and Lines Fluxes} \label{sec:moment_maps}

We generated velocity-integrated intensity, or ``zeroth moment," maps of line emission from the image cubes using \texttt{bettermoments} \citep{Teague18_bettermoments} with the same Keplerian masks employed during CLEANing and with no flux threshold for pixel inclusion to ensure accurate flux recovery. Rotation and peak intensity maps were generated using the `quadratic’ method of \texttt{bettermoments} with the full Planck function. The rotation maps used $5\sigma$-clipping to ensure only high-SNR line emission was used to estimate gas velocities. Appendix \ref{sec:app:peak_rotation} shows galleries of the resulting maps.

Continuum flux was extracted via the image directly (see next Section), while integrated line fluxes from spectra extracted using \texttt{GoFish} \citep{Teague19JOSS} and the same Keplerian masks used to generate the moment maps. Table \ref{tab:image_info} lists the integrated line fluxes and uncertainties.

\section{Results} \label{sec:results}

\subsection{Continuum Properties} \label{sec:continuum_results}

The left panel of Figure \ref{fig:mom0_gallery} shows the 0.9~mm continuum image of GoHam, which exhibits an elongated structure consistent with previous, lower-angular resolution images \citep{Bujarrabal08, Teague20_goham}. The dust is asymmetric along the major axis with an extended, narrow feature toward the south. No prominent asymmetries are seen along the minor axis and there is no clear evidence of rings or gaps at the angular resolution of our observations \citep[e.g.,][]{Villenave22}. 

To determine the geometry and continuum properties, we fitted the disk with a 2D Gaussian using the CASA task \texttt{imfit}. We measured a total flux of 0.881~$\pm$~0.003~Jy, which is modestly higher but generally consistent with the SMA observations of \citet{Bujarrabal09}. We derived a best-fit center position of R.A. (J2000)=18h09m13.413s, decl. (J2000)=$-$32$^{\circ}$10$^{\circ}$50.611$^{\prime \prime}$ and a position angle of 173$^{\circ}$.0$\pm$0$^{\circ}$.05. We adopted these parameters as the disk center and PA, respectively, for all subsequent analysis. The deconvolved full width at half maximum (FWHM) of the major and minor axes are 2\farcs360 $\pm$ 0\farcs007 and 0\farcs445 $\pm$ 0\farcs001, respectively, which implies an inclination of ${\approx}80^{\circ}$ for a geometrically-flat disk. Given that disks have finite vertical thickness, this inclination provides a lower limit and is consistent with the previous inference of $i{\approx}85^{\circ}$ from \citet{Bujarrabal08,Bujarrabal09}.

\begin{figure*}[t]
\centering
\includegraphics[width=0.875\linewidth]{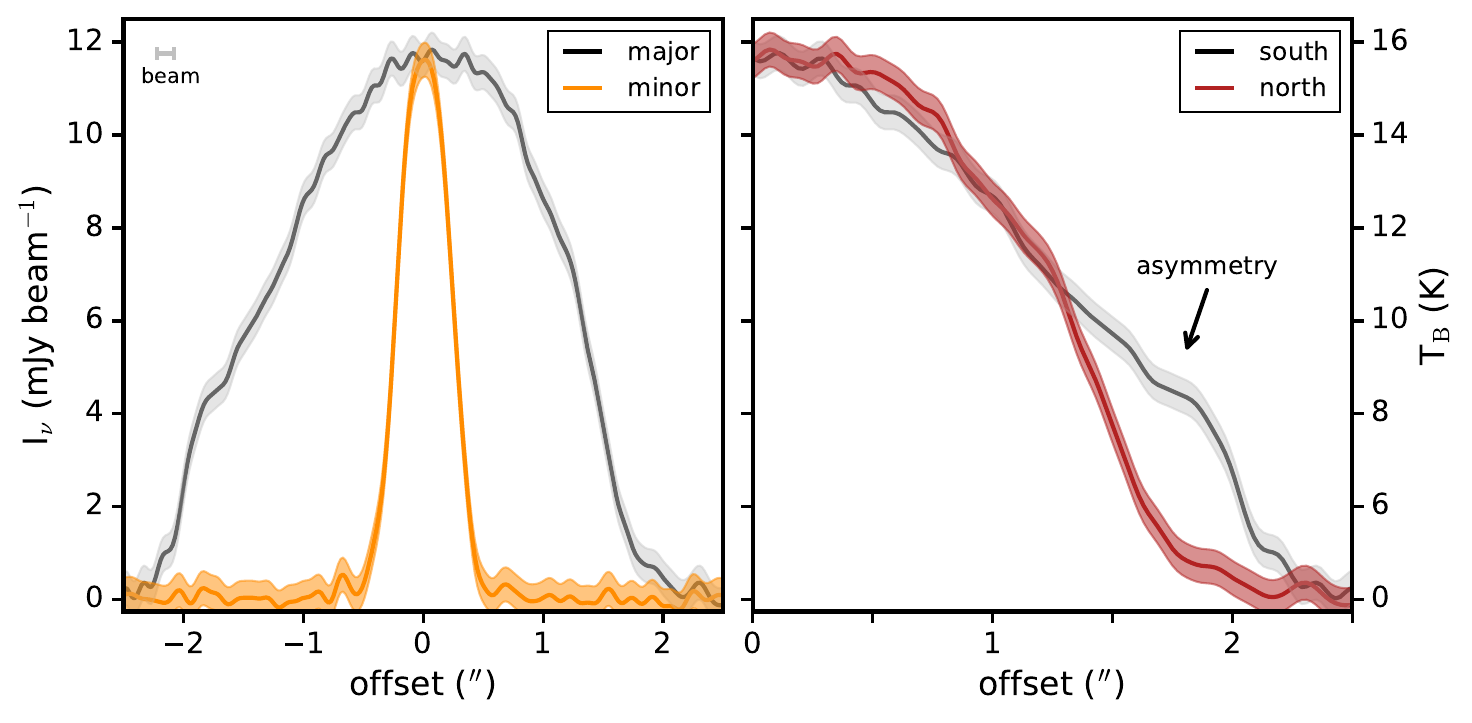}
\vspace{-10pt}
\caption{\textit{Left:} major (black) and minor (orange) axis cuts along the midplane of the 0.9~mm continuum. The positive direction for the major axis is along the northern part of the disk, while the positive direction for the minor axis is along the eastern part of the disk. The FWHM of the major axis of the synthesized beam is shown in the top left corner. \textit{Right:} Major axis cuts along the south (black) and north (red) directions illustrating the emission asymmetry. Brightness temperatures were computed using the full Planck function. The shaded regions indicate $\pm3\sigma$.}
\label{fig:cont_cuts}
\end{figure*}

Due to the north-south asymmetry, a simple 2D Gaussian fit results in significant disk-wide residuals (see Appendix \ref{sec:app:robust_continuum_imaging}). In an attempt to provide more robust geometric constraints, especially the inclination, we follow the continuum radiative transfer fitting approach outlined in \citet{Lin23} from which we adopted the same parameterized disk model and used RADMC-3D \citep{Dullemond12} to conduct ray-tracing without explicit heating and cooling calculations. Appendix \ref{sec:app:robust_continuum_imaging} provides a description of the disk model and best-fit parameters. For this approach, we fit the fiducial (\texttt{robust}=$-$0.5) and high-resolution  (\texttt{robust}=$-$2.0) images for the tightest possible geometric constraints. We obtain consistent results from both fits, and we report the high-resolution fit results here. The resulting parameters, together with the corresponding measurements from \texttt{imfit}, are summarized in Table \ref{tab:continuum_fit_rad_trans}.

The radiative-transfer fit yields a PA of $172^{\circ}.90\pm0^{\circ}.01$ and an inclination of $85^{\circ}.40\pm0^{\circ}.01$, which are consistent with the values inferred from \texttt{imfit} and previous SMA observations \citep{Bujarrabal08,Bujarrabal09}. The fitted stellar position, R.A. = 18h09m13.4093s and decl. = $-$32$^{\circ}$10$^{\prime}$50.560$^{\prime\prime}$, is also nearly identical to the \texttt{imfit} center. Thus, the two approaches provide consistent constraints on the bulk disk geometry despite the continuum asymmetry. The radiative-transfer model also recovers an integrated flux of 0.874 $\pm$ 0.009~Jy, consistent with the 0.881~$\pm$~0.003~Jy measured from \texttt{imfit}.

We emphasize that the uncertainties listed in Table \ref{tab:continuum_fit_rad_trans} only reflect the formal statistical uncertainties of fitting the adopted parametric model and do not capture systematic uncertainties associated with the model assumptions. In particular, the model assumes an axisymmetric disk and cannot reproduce the observed north-south asymmetry. Direct fitting of the interferometric visibilities would provide a more rigorous treatment of the spatially-resolved emission and would likely yield more robust constraints on the disk parameters. Such an analysis, including a more flexible treatment of the continuum asymmetry, is beyond the scope of this work. Nevertheless, the consistency between the two fitting approaches and with previous measurements gives us confidence in the bulk geometric parameters. We therefore adopt $i=85^{\circ}.40$ for all subsequent analysis.

\begin{deluxetable*}{llcc}
\tablecaption{Continuum Fitting Results \label{tab:continuum_fit_rad_trans}}
\tablewidth{0pt}
\tablehead{
\colhead{Description} & \colhead{Parameter} & \multicolumn2c{Value}  \\ \cline{3-4}
 & & \texttt{imfit} & Model Fit\tablenotemark{a} \\
}
\startdata
Int. Flux & S$_{\nu}$ & 0.881 $\pm$ 0.003 Jy & 0.874 $\pm$ 0.009 Jy \\
Inclination & i & ${>}$80$^{\circ}$ & 85$^{\circ}$.40$^{+0.01}_{-0.01}$\\
Position Angle & PA & 173$^{\circ}$.0$ \pm $0$^{\circ}$.05 & 172$^{\circ}$.90$^{+0.01}_{-0.01}$ \\
Disk Edge & R$_0$ & \ldots & 260.16$^{+1.35}_{-1.57}$~au \\
Temperature at R$_0$ & T$_0$ & \ldots & 9.47$^{+0.03}_{-0.02}$~K\\
Dust Scale Height at 100~au & H$_{100}$ & \ldots & 11.13$^{+0.01}_{-0.02}$~au \\
Characteristic Optical Depth & $\log_{10} \tau_0$ & \ldots & $-$0.268$^{+0.005}_{-0.004}$ \\
R.A. of Star & R.A. & 18h09m13.413s & 18h09m13.4093s \\
Decl. of Star & Decl. & $-$32$^{\circ}$10$^{\circ}$50.611$^{\prime \prime}$ & $-$32$^{\circ}$10$^{\circ}$50.560$^{\prime \prime}$ \\
\enddata
\tablenotemark{a}{As in \citet{Lin23}, we adopted a power-law brightness temperature prescription (with an exponent of $q=0.5$).}
\tablecomments{The reported errors represent statistical
uncertainties associated with fitting a functional form to
the extracted data points and do not include the (likely much larger) systematic uncertainties. Positional uncertainties are on the order of ${\approx}$0.1-0.4~mas.}
\end{deluxetable*}

Figure \ref{fig:cont_cuts} shows spatial cuts across the major and minor axes of the disk continuum emission. Both disk axes are spatially-resolved and the north-south asymmetry is clearly seen by reflecting the major axis profiles about the north-south direction. For radial separations within 1\farcs25, the north and south profiles are consistent, but beyond this point, the south shows an excess relative to the north and extends to a larger radius of ${\approx}$2\farcs2. 

From the measured integrated continuum flux, we can estimate a lower limit to the dust mass if we assume that all continuum emission is due to optically-thin thermal dust emission. Then, we can adopt the standard conversion used in disk settings \citep{Beckwith90}, i.e.,
\begin{equation}
    M_{\rm{dust}} = \dfrac{S_{\nu} d^2}{\kappa_{\nu} B_{\nu} (T)}
\end{equation}
where S$_{\nu}$ is the flux density, $\kappa_{\nu}$ is the dust mass opacity, d is the distance to the source, T is the dust temperature, and B$_{\nu}$ is the black body radiation computed from the Planck function. To be consistent with common assumptions adopted in disk surveys \citep[e.g.,][]{Ansdell16}, we take $\kappa = 2.3$~cm$^2$~g$^{-1}$ and T=20~K. This results in an M$_{\rm{dust}}$ of ${\approx}$180~M$_{\oplus}$ for the GoHam disk, which is significantly larger (${\approx}5\times$) than the mean dust masses of disks around Herbig stars \citep{Stapper22}, but even so, this still likely represents a significant underestimate given our assumption of optically-thin dust \citep[e.g.,][]{Tazzari21}. This further underscores the unusual scale, and high planet-forming potential, of the GoHam system.

\begin{figure*}
\centering
\includegraphics[width=.725\linewidth]{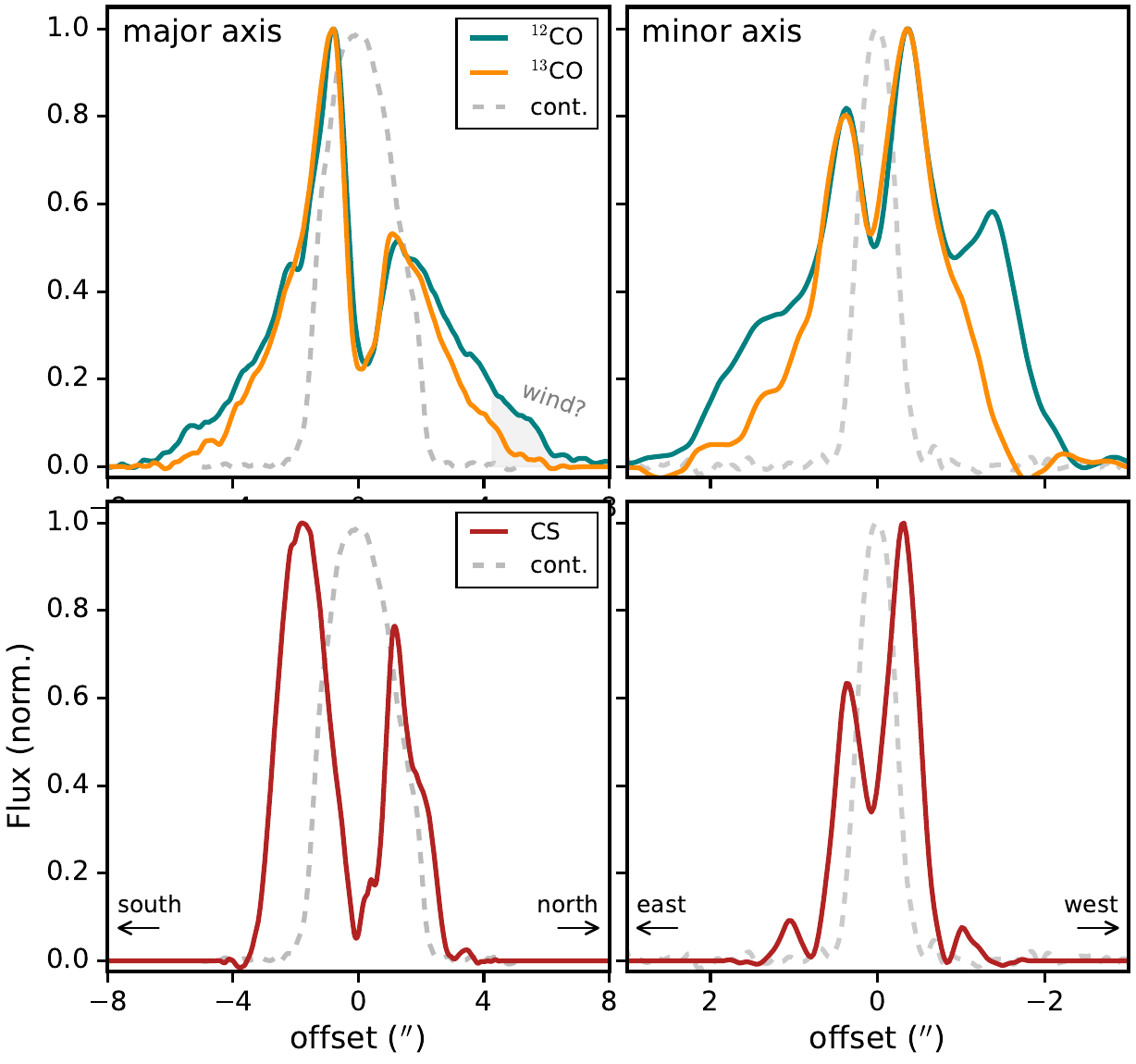}
\caption{Major (\textit{left column}) and minor (\textit{right column}) axis cuts of the $^{12}$CO (blue), $^{13}$CO (orange), and CS (red) integrated intensity maps versus mm dust continuum (dashed gray). All profiles have been smoothed for ease of comparison.}
\label{fig:line_cuts}
\end{figure*}

\begin{figure}
\centering
\includegraphics[width=\linewidth]{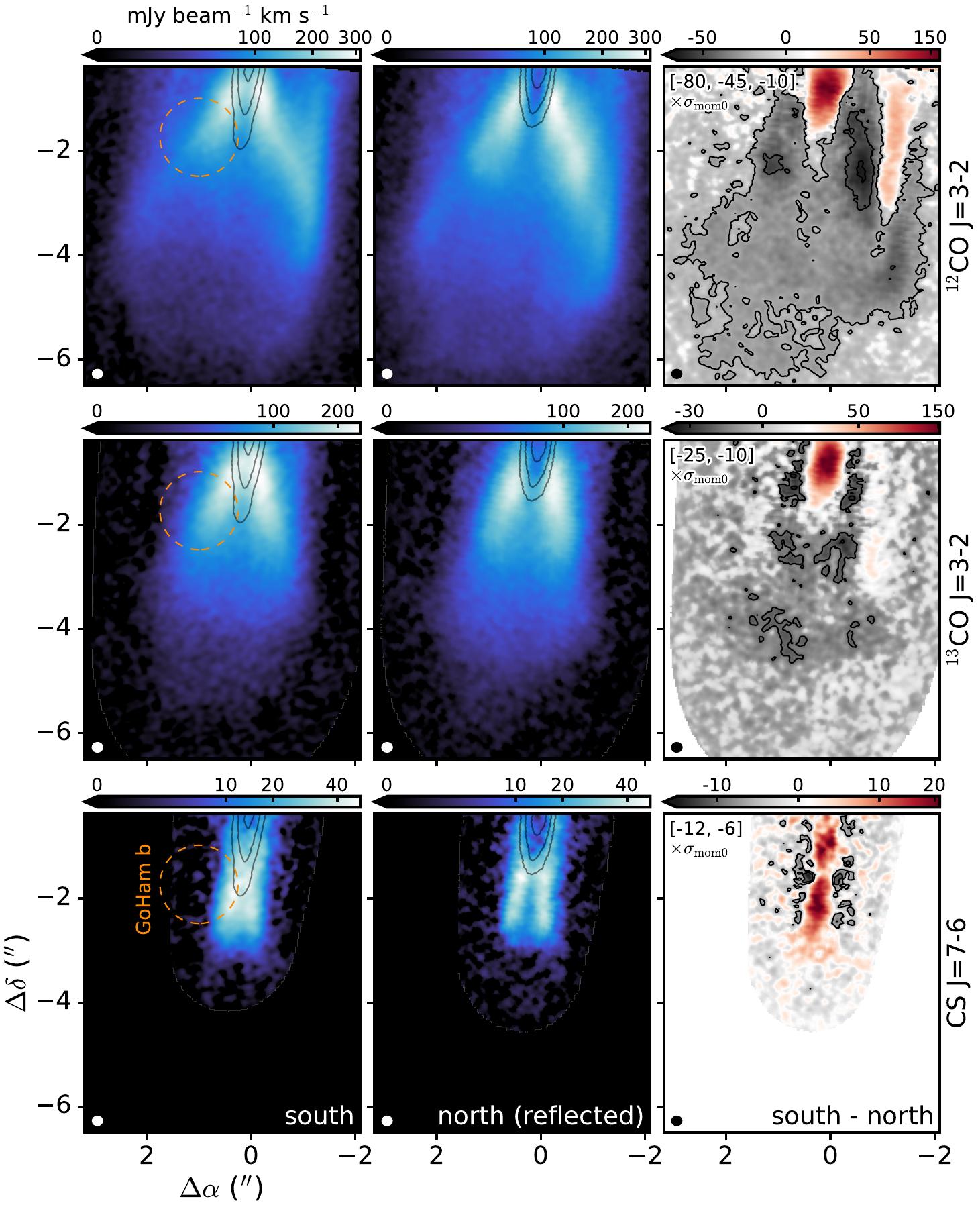}
\caption{Residual maps (\textit{right}) of $^{12}$CO, $^{13}$CO, and CS obtained by reflecting and subtracting the north (\textit{middle}) from south (\textit{left}) halves of the disk. Contours in the first two columns show the 0.9~mm continuum, while in the third column, they indicate negative residuals with significance levels in terms of the median zeroth moment map RMS value ($\sigma_{\rm{mom0}}$), as marked in the upper left corner of the residual panels. The approximate location of GoHam~b \citep{Bujarrabal09} is overlaid as a dotted orange circle.}
\label{fig:mom0_asym}
\end{figure}

\subsection{Morphology of Molecular Lines} \label{sec:line_results}

The right three panels of Figure \ref{fig:mom0_gallery} show zeroth moment maps of the $^{12}$CO, $^{13}$CO, and CS line emission. All lines show radially- and vertically-extended emission with emitting surfaces that are spatially-resolved and well-separated at nearly all disk positions. There is a clear vertical stratification pattern with $^{12}$CO emitting from the highest altitudes ($z/r \sim 0.5$-$0.6$), followed by $^{13}$CO at more moderate heights ($z/r \sim 0.4$), and then CS ($z/r \sim 0.15$), which lies closest to, but still above, the disk midplane (as traced by the mm dust). These characteristic heights are approximate visual estimates based on the apparent projected separation of the emission surfaces from the midplane and the expected geometry of the GoHam disk. A detailed measurement of the emission surfaces will be presented in a forthcoming analysis. Overall, this structure is consistent with previous observations of other edge-on and inclined disks at high spatial resolutions \citep[e.g.,][]{Dutrey17, pinte18, Law21, Galloway25, Dutrey25}.

All lines also show a central emission deficit near the location of the disk midplane, which manifests as a dark lane in Figure \ref{fig:mom0_gallery}. For $^{12}$CO, this dark lane is spatially coincident with the region occupied by the mm continuum, while the correspondence is also present, but slightly less clear, in $^{13}$CO and CS. This effect can also be seen in the peak intensity maps (Appendix~\ref{sec:app:peak_rotation}). The dark lane likely results from a combination of three effects: molecular freeze-out in the coldest regions near the midplane \citep{Podio20}, obscuration of line emission by optically-thick dust along the line of sight \citep{Bosman21}, and continuum over-subtraction during the imaging process \citep{Boehler17,Weaver18,Rosotti21}. To assess the latter effect, we examined the non-continuum-subtracted line images and find that the dark lane remains present, although with reduced contrast, indicating that it is not solely an imaging artifact but that continuum subtraction enhances its depth. Moreover, molecular gas remains warm at radii beyond the mm continuum, particularly for optically-thick $^{12}$CO (see the peak intensity maps in Appendix~\ref{sec:app:peak_rotation}). Thus, freeze-out may contribute to the central deficit near the cold midplane, but a purely radial temperature-dependent freeze-out boundary cannot fully explain the observed morphology or its close correspondence with the continuum emission.

To better investigate the line emission morphology, Figure \ref{fig:line_cuts} shows cuts along the major and minor axes of the disk in the $^{12}$CO, $^{13}$CO, and CS zeroth moment maps. For each of these profiles, we also applied a \texttt{savgol\_filter} in \texttt{scipy} \citep{Virtanen_etal_2020} to increase profile smoothness for ease of comparison and ensure that we do not over-interpret spurious small-scale structures with low-SNRs.

\textit{Major Axis:} $^{12}$CO and $^{13}$CO show nearly identical major axis profiles extending to radial distances of $\pm$6$^{\prime \prime}$, while the CS profile is considerably more compact with a radius of $\pm$3$^{\prime \prime}$. All lines show a central dip in emission that approximately corresponds to the mm dust peak. There is also a southern brightness asymmetry at $\pm$1-2$^{\prime \prime}$ along the midplane that ranges from a factor of two in $^{12}$CO and $^{13}$CO to a more modest asymmetry of ${\approx}$20\% in CS. This asymmetry does not occur at the same radius with CS, showing enhancements at slightly larger radii than in $^{12}$CO and $^{13}$CO. In addition, there is a secondary asymmetry at larger radii in both $^{12}$CO and $^{13}$CO, which shows more extended, brighter emission toward the northern part of the disk versus the southern region. Although more apparent in the north and south major-axis profiles (Figure \ref{fig:line_cuts}), this asymmetry remains subtle compared to the stronger asymmetries described above and may be related to the potential wind signature discussed in Section \ref{sec:wind_CO}.

\textit{Minor Axis:} All three lines show two minor axis peaks at $\pm$0\farcs3-0\farcs4, with a central depression that matches the continuum profiles. These peaks correspond to the locations where the molecular emitting surfaces become visible along the minor-axis cut, as the line-of-sight dust opacity decreases away from the midplane, allowing emission from the warm molecular layers to emerge, and the temperature becomes sufficiently high to prevent substantial freeze-out onto grain surfaces. The fainter peak toward the eastern portion of the disk is consistent with this side of the disk being tilted toward the observer, which is confirmed in the peak intensity maps (Appendix \ref{sec:app:peak_rotation}). $^{12}$CO also shows two extra features, one sharper peak at $-$1\farcs4 and another more extended shoulder from $+$1-2$^{\prime \prime}$, which manifest as vertical bands in the zeroth moment map (Figure \ref{fig:mom0_gallery}). Both features may arise from changes in the line-of-sight geometry and optical depth at these projected offsets, which allow emission from the far-side molecular layer to contribute to the observed emission \citep[e.g.,][]{Dutrey17, Flores21}. The relative brightness differences and slopes of these features are set by the east-west tilt of the disk orientation. The fact that we do not see these features in non-$^{12}$CO lines may point to the unique sensitivity of $^{12}$CO in tracing cold, faint, and low-density disk gas.

To better map the spatial distribution of these asymmetries, we take the northern half of the disk and subtract it from the southern half after reflecting about the disk center. Figure \ref{fig:mom0_asym} shows the resulting residual maps for $^{12}$CO, $^{13}$CO, and CS. Excess midplane emission in the southern part of the disk is present in all lines, as reflected in the positive residuals. Notably, the structure of these residuals is different, with CS having excess emission at larger radii than either $^{12}$CO or $^{13}$CO but with a \added{narrower overall East-West width.} The excess emission in $^{12}$CO and $^{13}$CO is well-aligned with the morphology of the mm dust continuum, as shown in Figure \ref{fig:mom0_asym}. In particular, the northern side of the continuum disk is brighter but less extended, while the southern side is fainter but extends to larger radii. The stronger $^{12}$CO and $^{13}$CO deficits toward the northern side coincide spatially with the brighter continuum emission, consistent with greater dust optical depth preferentially obscuring the molecular emission there. Conversely, the more extended but fainter southern continuum is associated with a weaker suppression of the $^{12}$CO and $^{13}$CO emission. This correspondence suggests that dust opacity contributes significantly to the observed midplane asymmetries in $^{12}$CO, $^{13}$CO, and the inner part of CS. However, the CS midplane asymmetry seen at large radii cannot be explained by continuum optical depth.

Vertical bands in the western part of the disk are also apparent in $^{12}$CO, and to a lesser degree in $^{13}$CO. These approximately vertical features likely reflect the far-side emission contribution seen in the minor-axis cuts, with their unequal north-south strengths arising from the three-dimensional geometry of the nearly edge-on disk, which causes the line of sight to intersect different portions of the far-side emitting layer on either side of the disk. The appearance of this banding structure in $^{13}$CO, but not in the profiles in Figure \ref{fig:line_cuts}, is likely due to its relatively low-SNR (i.e., it is not clearly detected in the narrow intensity cuts shown in Figure \ref{fig:line_cuts}). Prominent (${\sim}10\sigma$) negative $^{12}$CO residuals, extending to radii of ${\gtrsim}800$~au, are associated with the diffuse emission present in the northern part of the disk, which we discuss in the following Section. These residuals may also partly reflect differences in the vertical location of the $^{12}$CO emitting layer between the northern and southern sides of the disk.

In Figure \ref{fig:mom0_asym}, we also mark the location of GoHam~b from \citet{Bujarrabal09}, which does not exhibit any obvious residual signal in either $^{12}$CO or $^{13}$CO. However, given the complex and widespread residual structure across the disk, this is not unexpected, and a definitive assessment of the potential $^{13}$CO gas over-density associated with GoHam~b will require detailed radiative transfer modeling.

\subsection{Possible Disk Wind Signatures in $^{12}$CO and $^{13}$CO} \label{sec:wind_CO}

\begin{figure*}
\centering
\includegraphics[width=\linewidth]{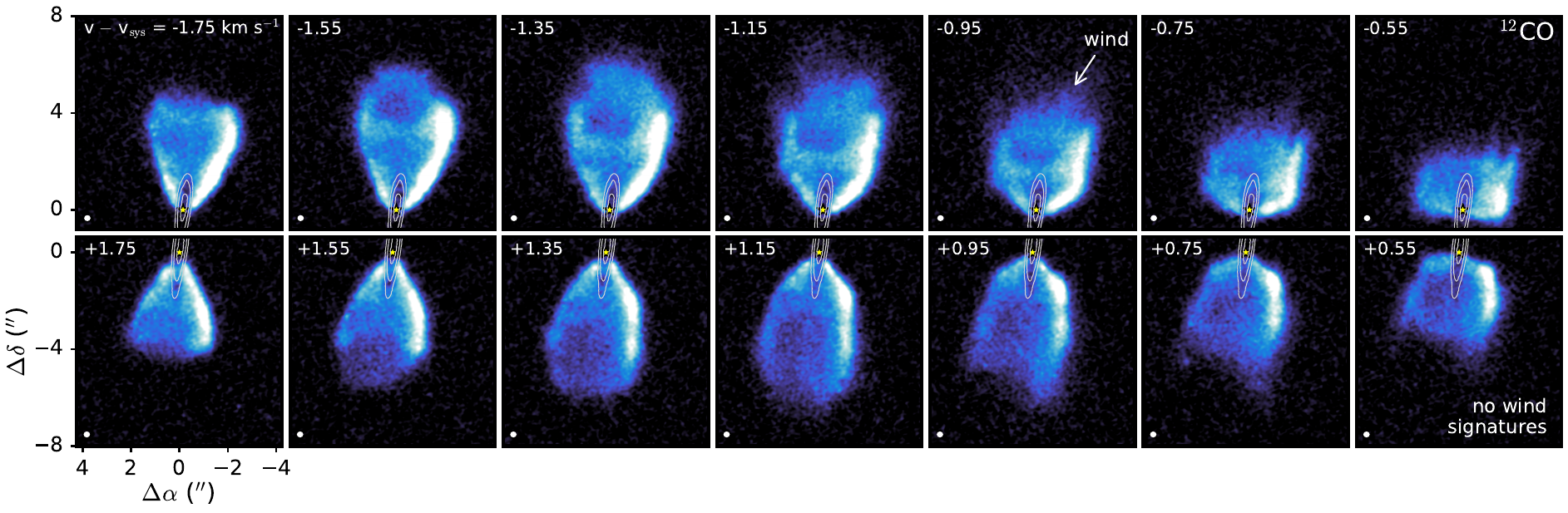}
\includegraphics[width=\linewidth]{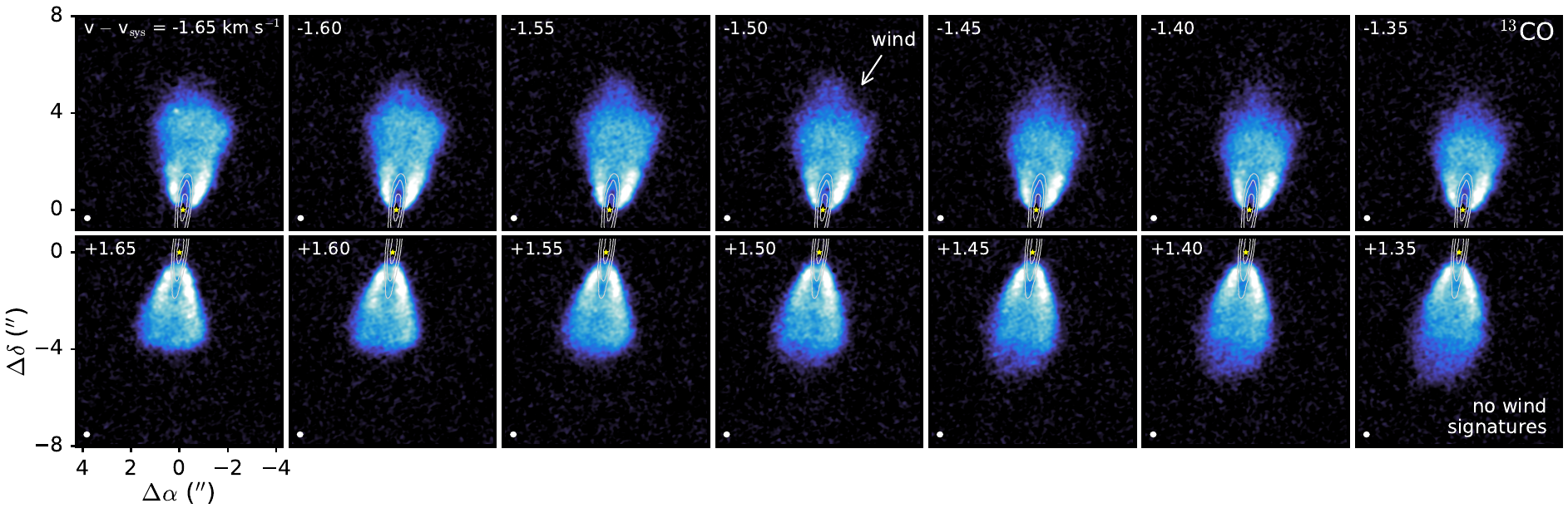}
\caption{Representative channels of $^{12}$CO J=3--2 (\textit{top}) and $^{13}$CO J=3--2 (\textit{bottom}) showing extended, wind-like emission (marked by arrows) in the north of the GoHam disk. No corresponding extended emission is seen in the southern part of the disk at the same relative velocities in either molecule. Channels are shown in pairs at approximately equal and opposite velocity offsets from the systemic velocity (v$_{\rm sys}\approx2.8$~km~s$^{-1}$). For visual clarity, only a subset of channels is shown, with $^{13}$CO displayed at a finer velocity sampling and narrower total velocity range to highlight its more subtle wind-like features. The synthesized beam is shown in the lower left corner of each panel.}
\label{fig:CO_wind}
\end{figure*}

There is evidence for extended $^{12}$CO and $^{13}$CO emission at large radii toward the northern regions of GoHam that deviates from the expected Keplerian disk emission and may be associated with outflowing gas or a disk wind. This emission is apparent in the asymmetric major-axis line profiles in Figure \ref{fig:line_cuts} and the negative residuals in Figure \ref{fig:mom0_asym}. To illustrate these features, Figure \ref{fig:CO_wind} shows representative channel maps highlighting this extended emission. The extended $^{12}$CO emission is detected over a V$_{\rm{LSR}}$ range of ${\approx}$1-2.5~km~s$^{-1}$, while $^{13}$CO traces a narrower, but overlapping, velocity range of ${\approx}$1-1.5~km~s$^{-1}$. 

The $^{12}$CO features take the form of a wide-angle, conical-like structure, most apparent from a V$_{\rm{LSR}}$ of 1.5-2~km~s$^{-1}$, while closer to the systemic velocity (v$_{\rm{sys}}\approx$ 2.8~km~s$^{-1}$), they consist of several wisp-like, linear bands originating from several different disk heights. The $^{13}$CO, although more subtle, traces an extended, but narrow, protrusion beyond the Keplerian disk extent. The redshifted velocities of these features suggest that material entrained in this potential wind is moving away from the observer, but due to projection effects and in the absence of a dedicated kinematic model, it is not clear if it arises from the front or back (or both) sides of the disk. Since the MRS of our observations is 5\farcs7, these structures may only trace the inner portion of a larger-scale outflow. Follow-up observations optimized for larger angular scales will be key to map the full spatial extent of this extended component.

The observed CO structure in GoHam has several similarities to that of IM~Lup system, a well-studied example of a nearby disk undergoing external photoevaporation. IM~Lup hosts a large gas disk with $^{12}$CO emission extending to radii of ${\sim}$1000~au \citep{Cleeves_IMLup_16, Law21_MAPS3}, spiral structure in its millimeter dust indicative of a massive and potentially gravitationally unstable disk \citep{Huang18_spiral_imlup}, and diffuse, non-Keplerian $^{12}$CO emission that has been successfully reproduced by models of external photoevaporation driven by a relatively weak (G$_0$=4) ambient external radiation field \citep{Haworth17}. These similarities motivate external photoevaporation as one possible explanation for the extended CO emission in GoHam. The revised distance places GoHam in the outskirts of Sco-Cen (Appendix \ref{sec:app:revised_distance}), the nearest OB association to the Sun, which contains numerous massive stars capable of producing an enhanced external FUV radiation field \citep{Preibisch08,Wright18,Anania25}. However, the local irradiation environment is poorly constrained, because three-dimensional source distances and intervening extinctions are uncertain. Recent studies show that even modest fields of G$_0\sim$ 2-12 can influence disk evolution \citep{Anania25_AGEPRO}, so external photoevaporation cannot be excluded without a robust estimate of the local radiation field. 

Alternatively, the extended CO emission in GoHam may trace a magnetohydrodynamic (MHD)-driven wind \citep[e.g.,][]{Bai16, Bethune17, Louvet18, deValon22, Pascucci25}. The observed morphology in Figure \ref{fig:CO_wind} is not uniquely diagnostic of external photoevaporation, and some aspects of the emission, including its large vertical extent, wide-angle morphology, and apparent origin from multiple disk heights, are also broadly consistent with magnetically-launched outflows. In particular, the diffuse CO emission in GoHam resembles the extended $^{12}$CO J=2--1 wind-like feature reported in the edge-on Tau~042021 disk \citep{Duchene24}. Observations with HST and JWST \citep{Duchene14, Duchene24, Arulanantham24, Pascucci25} revealed both a collimated atomic jet ([Ne~II], [Fe~II]) and wide-angle H$_2$ molecular wind, a combination that strongly favors an MHD-wind origin. While our data do not distinguish between photoevaporative and MHD-driven scenarios in GoHam, the similarity to Tau~042021 highlights that multiple wind-launching mechanisms can produce comparable large-scale CO morphologies.

Overall, we emphasize that our observations do not provide a definitive identification of a disk wind, but rather reveal an extended, non-Keplerian CO component for which this is one possible interpretation. Dedicated kinematic modeling will be required to constrain its launching geometry and distinguish between possible wind scenarios. Deep IFU spectro-imaging observations (i.e., with JWST NIRSpec, MIRI) capable of resolving the kinematics and excitation structure of the extended gas will likewise be critical for constraining its origin and physical properties.

\subsection{Dynamical Stellar Mass Estimates} \label{sec:m_dyn}

The Keplerian rotation of the gas around the central star in the GoHam system (Appendix \ref{sec:app:peak_rotation}) offers a powerful opportunity to directly measure the dynamical stellar mass. We make use of a position-velocity (PV) diagram-based method, as is commonly done for edge-on protostellar and protoplanetary systems \citep[e.g.,][]{Seifried16, Flores23, vanHoff23, Lin23, Phuong25}. To do so, we first generated PV diagrams of $^{12}$CO, $^{13}$CO, and CS along the major axis of the disk using the \texttt{impv} task in CASA. We then extracted rotation curves from these PV diagrams using the \texttt{pvanalysis} module in the Spectral Line Analysis/Modeling (SLAM) code \citep{Aso23, Aso24}. While a full description of SLAM is outlined in \citet{Ohashi23}, we briefly describe the procedure we adopted below.

We used both the ridge and edge methods to determine points for subsequent rotation curve fitting. The ridge method identifies the characteristic velocity at each spatial position by fitting a 1D Gaussian to the emission profile along the velocity axis of the PV diagram, i.e., tracing the centroid near the emission peak. In contrast, the edge method locates the outer boundary of the emission by identifying the $5\sigma$ contour in the PV diagram, which samples the highest-velocity emission at small radii and the most spatially-extended emission at lower velocities. We then fit the resulting positions and velocities $(r,v)$ from each method independently with the following function:

\begin{equation}
    v = \rm{sign}(r) \, v_b \left( \dfrac{|r|}{r_b} \right)^{-p(r)} + v_{\rm{sys}},
\end{equation}

where $r_b$ is the characteristic radius, $v_b$ is the rotational velocity at $r_b$, and 

\begin{equation}
    p(r) =
    \begin{cases} p_{\rm in}, & |r| < r_b,\\
    p_{\rm in} + dp, & |r| \geq r_b,
    \end{cases}
\end{equation}

\noindent where $p_{\rm in}$ is the inner power-law index of the rotation profile, and $dp$ is the change in the power-law index beyond $r_b$. Due to the presence of known asymmetries in the line emission, we opted to fix $p_{\rm{in}}=0.5$ to ensure that we only consider the bulk Keplerian rotation pattern. Likewise, we allowed $dp$ to be a free parameter for the $^{12}$CO and $^{13}$CO fits in light of the wind-like signatures seen in these lines, but fixed $dp=0$ for CS. Then, we can infer the stellar mass by assuming Keplerian rotation via $v_b = \sqrt{G M_*/r_b} \sin i$.

\begin{figure*}
\centering
\includegraphics[width=\linewidth]{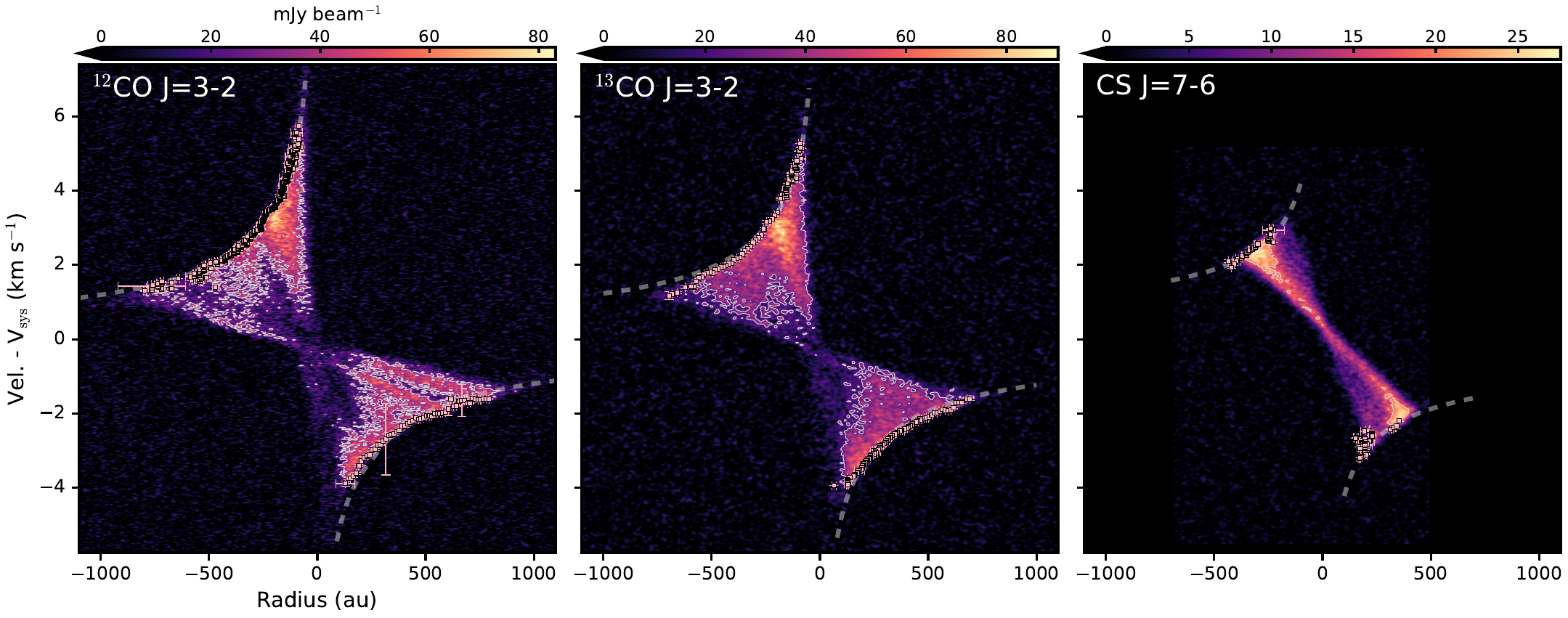}
\caption{Position-velocity diagrams of $^{12}$CO (\textit{left}), $^{13}$CO (\textit{middle}), and CS (\textit{right}) showing the data points (squares) and best-fit rotation curve (dashed curve) derived using the edge method. Contour levels in white indicate 5$\sigma$. The PV diagrams have been shifted by the systemic velocity (v$_{\rm{sys}}\approx$ 2.8~km~s$^{-1}$).}
\label{fig:PV_fits}
\end{figure*}

Table \ref{tab:dyn_stellar_mass} lists the best-fit values, and Figure \ref{fig:PV_fits} shows the resulting Keplerian rotation curves overlaid on the PV diagrams from which they were determined. The ridge and edge methods generally recover the same bulk Keplerian rotation pattern, although the selected points, and hence characteristic velocities, differ between methods and tracers. The lowest stellar masses are derived using CS, with the ridge method resulting in a mass potentially as low as ${\approx}$1.7~M$_{\odot}$, while $^{12}$CO gives the highest inferred masses, where the edge method suggests a stellar mass of ${\approx}$3.0~M$_{\odot}$. However, the ridge method is known to systematically underestimate uncertainties, while the edge method tends to produce overestimates of stellar mass \citep{Aso15, Maret20}. Variations in the inferred stellar masses between molecules may also reflect varying optical depths and the fact that each species traces a different vertical layer. These effects can shift the characteristic velocities measured in the PV diagrams and thus introduce systematic differences in the inferred stellar mass, even when the underlying rotation is governed by the same gravitational potential. This is particularly relevant for an edge-on disk, where the line of sight intersects a broad range of radii and heights, although the geometry also provides a favorable view of the rotational velocity \citep[e.g.,][]{Seifried16,Maret20}. Given the range of estimated masses, we adopt the mean value across methods and lines, resulting in a dynamical mass of M$_*$ = 2.2~$\pm$~0.5~M$_{\odot}$ and systemic velocity of v$_{\rm{sys}}$ = 2.75~$\pm$~0.06~km~s$^{-1}$.

\begin{deluxetable*}{lccccc}
\tablecaption{Dynamical Stellar Mass of GoHam Inferred from PV Diagram Fits\label{tab:dyn_stellar_mass}}
\tablewidth{0pt}
\tablehead{
\colhead{Method} & \colhead{$r_b$} & \colhead{$v_b$} & \colhead{$dp$}  & \colhead{M$_*$} & \colhead{v$_{\rm{sys}}$} \\ 
\colhead{} & \colhead{(au)} & \colhead{(km~s$^{-1}$)} & \colhead{} & \colhead{(M$_{\odot}$)} & \colhead{(km~s$^{-1}$)} 
}
\startdata
\textbf{$^{12}$CO} & \\
Edge              & 251.54 $\pm$ 2.24 & 3.255 $\pm$ 0.016 & 0.226 $\pm$ 0.002 & 3.023 $\pm$ 0.041 & 2.723 $\pm$ 0.002    \\
Ridge             & 213.03 $\pm$ 0.04 & 2.852 $\pm$ 0.002 & 0.189 $\pm$ 0.002 & 1.965 $\pm$ 0.003 & 2.752 $\pm$ 0.002  \\ \\
\textbf{$^{13}$CO} & \\
Edge              & 263.65 $\pm$ 1.48 & 2.945 $\pm$ 0.010 & 0.154 $\pm$ 0.004 & 2.594 $\pm$ 0.022 & 2.700 $\pm$ 0.002   \\
Ridge             & 213.02 $\pm$ 0.02 & 2.653 $\pm$ 0.002 & 0.146 $\pm$ 0.002 & 1.701 $\pm$ 0.003 & 2.685 $\pm$ 0.002   \\ \\
\textbf{CS}       &&& \\
Edge              & 288.22 $\pm$ 0.75 & 2.486 $\pm$ 0.020 & [0.0] & 2.020 $\pm$ 0.005 & 2.838 $\pm$ 0.003 \\
Ridge             & 239.30 $\pm$ 0.50 & 2.486 $\pm$ 0.020  & [0.0] & 1.677 $\pm$ 0.004 & 2.815 $\pm$ 0.003  \\
\enddata
\tablecomments{The ridge method was derived using a threshold of 5$\sigma$ for all lines. In all fits, we fixed p$_{\rm{in}}$=0.5 to ensure that we only consider Keplerian rotation to retain the interpretation that M$_*$ is the central stellar mass. $dp$ was kept as a free parameter for $^{12}$CO and $^{13}$CO but fixed to $dp=0$ for CS, as indicated in brackets.}
\end{deluxetable*}

\subsection{Detection of Asymmetric SO Emission} \label{sec:detection_of_SO}

\begin{figure}
\centering
\includegraphics[width=\linewidth]{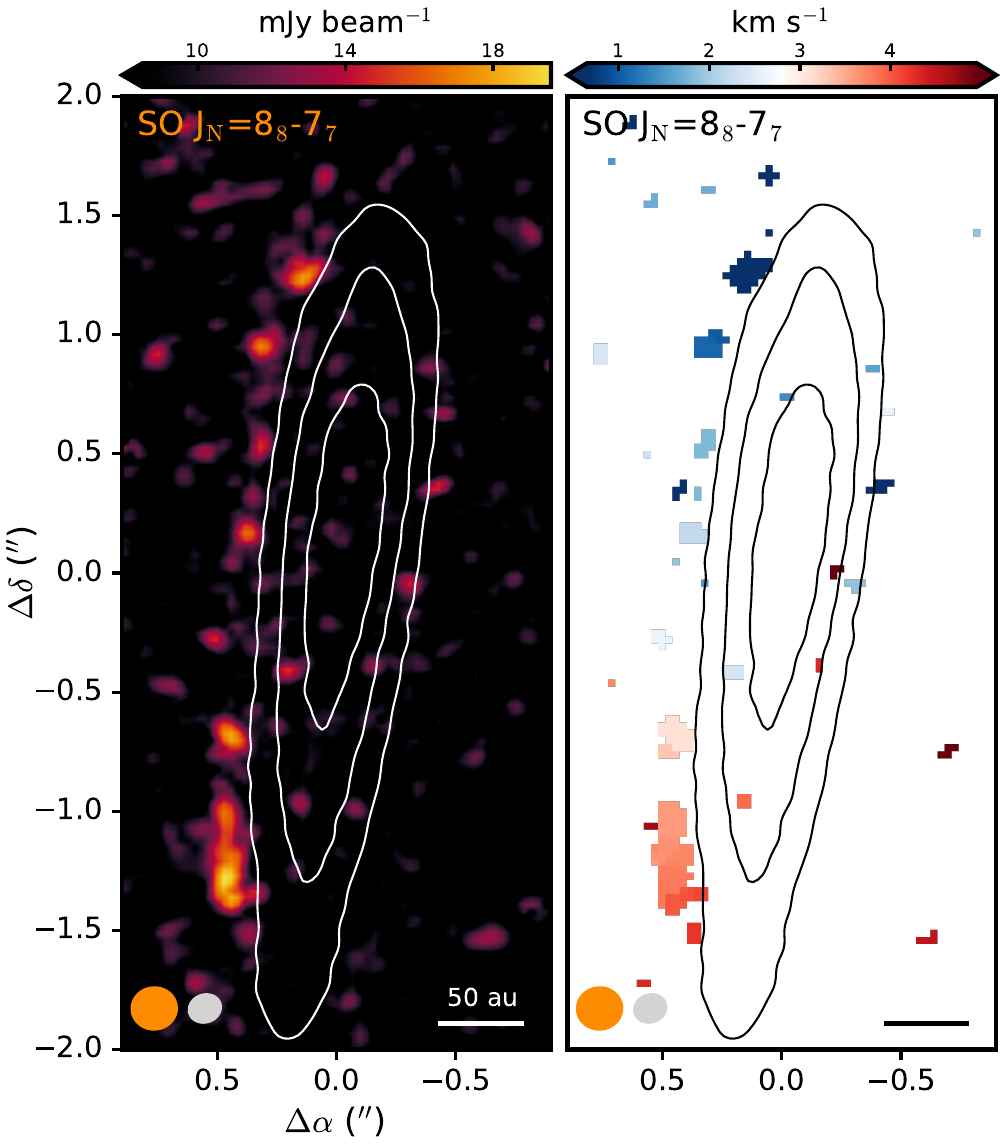}
\caption{Peak intensity (\textit{left}) and rotation (\textit{right}) maps of SO J$_{\rm{N}}$=8$_8$--7$_7$ in the GoHam disk. Contours correspond to the 0.9~mm continuum and show RMS$\times$[25, 55, 85]. Synthesized beams of the SO (orange) and continuum (light gray) images are shown in the lower left corner of each panel.}
\label{fig:SO_detection}
\end{figure}

In addition to $^{12}$CO, $^{13}$CO, and CS, we detected emission from SO J$_{\rm{N}}$=8$_8$--7$_7$, but it does not show the same morphology. The left panel of Figure \ref{fig:SO_detection} shows the peak intensity of the detected SO line. Strikingly, the SO emission is located only on the eastern side (i.e., the near side) of the disk and lies at the edge of the dust continuum. Its emission morphology takes the form of a broad arc-like feature comprising multiple, compact emission components rather than a fully-continuous distribution. The gaps between these features are larger than our beam size, so we do not believe these to be artifacts, but we cannot rule out the presence of a more continuous band of fainter SO emission. Even at the high-angular resolution of our observations (${\approx}$0\farcs2), the vertical extent of this SO emission is unresolved, implying a narrow height of ${<}$30~au.

In addition, there is a north-to-south brightness asymmetry with the southernmost region at $r{\approx}$1\farcs3-1\farcs4 (${\approx}$180-195~au) being the brightest and showing the largest spatially-continuous region of SO emission. The peak intensity measured within the southernmost SO component is around 20\% higher than the peak intensity of the northernmost portion of SO emission. Using a one-beam aperture centered on each of these regions, the integrated flux of the southern component is nearly one-third higher, reflecting its larger spatial extent. The SO emission appears to broadly trace the eastern edge of the mm dust continuum, with the southern clump potentially being spatially associated with the southern mm dust asymmetry.

Despite the unusual spatial distribution, the SO kinematics are consistent with Keplerian rotation as illustrated in the right panel of Figure \ref{fig:SO_detection}. Given that the southern bright region of SO emission is close to the predicted location of GoHam~b, one possibility for the origin of this emission, which we return to in the discussion (Section \ref{sec:SO_emission}), is from local heating driving the sublimation of sulfur-bearing ices, which, in turn, generates gas-phase SO that subsequently freezes out as it rotates away from its heating source. 

\section{Discussion} \label{sec:discussion}

Our ALMA observations reveal a substantially more complex disk than was apparent in existing lower-resolution (${\sim}$1$^{\prime \prime}$) SMA data \citep{Bujarrabal08, Bujarrabal09, Teague20_goham, Cusson26}, with rich spatial and kinematic structure, which is especially evident in $^{12}$CO and $^{13}$CO. This newly-revealed complexity complicates the identification of localized $^{13}$CO emission in excess of a smooth Keplerian disk model, as was previously used to identify GoHam~b \citep{Bujarrabal09, Teague20_goham}. The spatial and kinematic asymmetries seen with ALMA indicate that such residuals cannot be uniquely attributed to a localized $^{13}$CO gas overdensity without first accounting for the underlying disk structure, e.g., through detailed radiative transfer modeling beyond the scope of this work. We therefore do not attempt to confirm or refute the previously-proposed GoHam~b interpretation with the present data. Here, we instead focus on the overall disk structure to provide context for the newly-detected SO emission, which emerges as a clear tracer of localized heating and chemical processing associated with a dynamical perturbation. A detailed tomographic and kinematic analysis of the $^{12}$CO, $^{13}$CO, and CS emission, including the region around GoHam~b, will be presented in the forthcoming work of \citet{jensen_goham}.

\subsection{Gas and Dust Structure of GoHam} \label{sec:height_compar}

\subsubsection{Scattered Light Morphology} \label{sec:scattered_light_morph}

Before discussing the origin of the SO emission, we first examine the broader dust and gas structure of the GoHam disk. While the ALMA continuum traces millimeter-sized particles, optical and near-infrared scattered-light imaging is sensitive to micron-sized grains that remain more closely coupled to the gas. For this comparison, we make use of existing HST images from program 9315 (PI: K. Noll) that observed GoHam on 2002-02-22 using WFPC2 and the F450W, F555W, and F675W filters with a total integration time of 1940~s \citep{Wood08}. We adopt the combined, multi-color image obtained from the Hubble Legacy Archive. 

As shown in Figure \ref{fig:NIR_vs_lines}, GoHam has the typical structure of an edge-on disk viewed at optical/NIR wavelengths, namely a central dark lane surrounded by two lobes of extended, scattered light emission, as the light from the central star is blocked by the disk midplane \citep[e.g.,][]{Burrows96, Wolf03, Perrin06, Villenave20, Duchene24, Tazaki25}. While a detailed analysis of the small dust properties is beyond the scope of this work, we note a few salient morphological features apparent in Figure \ref{fig:NIR_vs_lines}.

The western half of the disk is ${\approx}$10--15\% brighter than the eastern lobe, while filament-like emission extends to large radii on both north and south sides of the disk. These features may trace small dust grains entrained in gas flows along the disk surface \citep[e.g.,][]{Franz20, Rodenkirch22, Duchene24}. These filaments reach radial distances of ${\sim}$6$^{\prime \prime}$, comparable to the extent of the $^{12}$CO emission (see Figure \ref{fig:line_cuts}). Similar, tail-like structures have been seen in other edge-on sources, including the disks around IRAS~23077+6707 \citep[`Dracula’s Chivito;'][]{Berghea24, Monsch25} and HH~30 \citep{Tazaki25}. However, unlike those systems, which exhibit strongly one-sided scattered-light structures, the GoHam filaments are present on both sides of the disk. This symmetry argues against a localized origin such as late infall, which is expected to produce asymmetric structures as material is accreted from a particular direction \citep[e.g.,][]{Akiyama19,Ginski21,Garufi22}. Instead, the comparable radial extents of the scattered-light filaments and $^{12}$CO emission suggest that the filaments are associated with the extended gas disk rather than a remnant envelope or captured cloud material. We also note that the non-Keplerian $^{12}$CO and $^{13}$CO emission discussed in Section~\ref{sec:wind_CO} is detected only on the northern side of the disk, whereas the scattered-light filaments extend in both directions. Thus, the filaments do not appear to trace the wind-like CO emission alone, suggesting that multiple dynamical processes may contribute to sustaining this extended material.

Small-scale, wispy structures are present at high elevations along the eastern side of the disk. These structures are similar, but less prominent, to those seen in Dracula’s Chivito, which were ascribed to ongoing dynamical activity in that disk \citep{Monsch25}. Unlike Dracula's Chivito, however, the wisps in GoHam are confined to a single disk surface rather than appearing on both sides of the disk, suggesting a more localized origin. The eastern lobe also hosts a brightness discontinuity near the southern edge of the millimeter dust disk, potentially tracing a gap in the small-grain distribution. This feature is located near GoHam~b (Figure \ref{fig:goham_b_vs_dust}) and may be associated with a localized perturbation in the southern half of the disk. While we note that such a perturbation need not produce identical signatures on the two disk surfaces, the origin of this one-sided morphology remains uncertain (see Section \ref{sec:vortex_continuum}).

\subsubsection{HST Versus ALMA Gas and Dust Heights}

\begin{figure*}
\centering
\includegraphics[width=\linewidth]{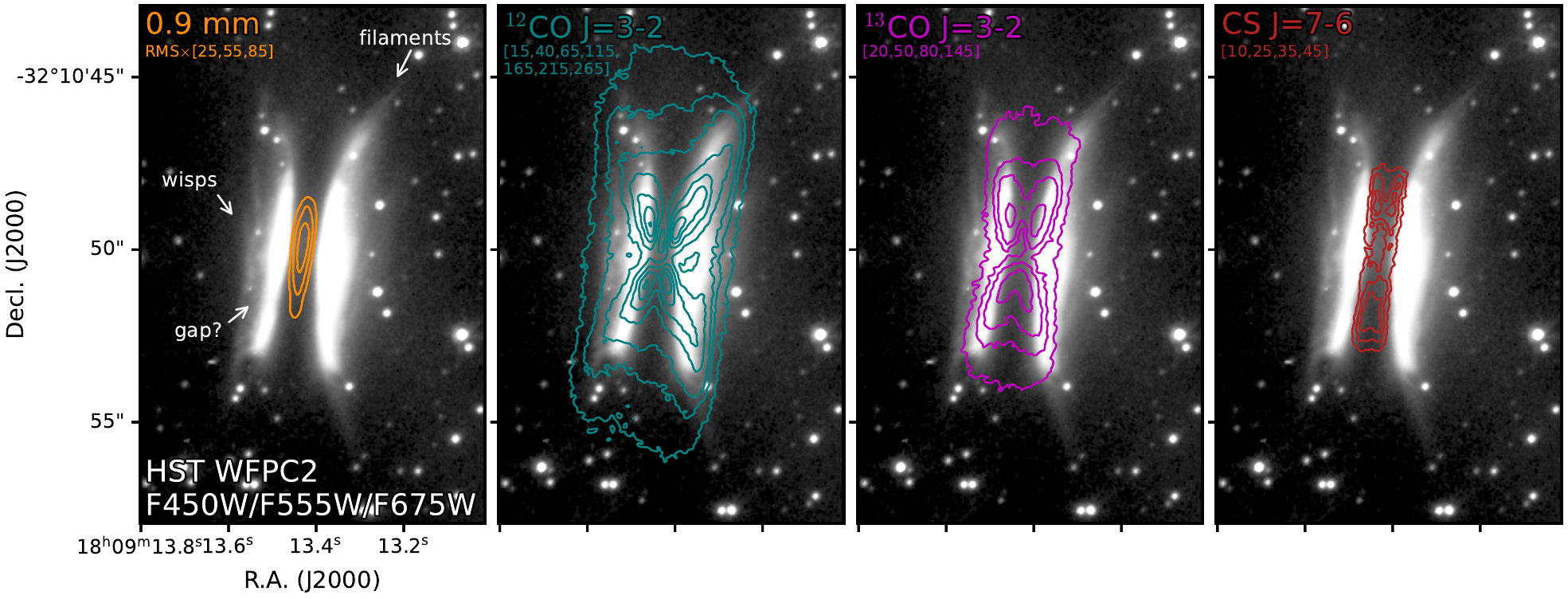}
\caption{HST WFPC2 composite image of GoHam sensitive to scattered light from small dust grains in the disk upper layers using the F450W, F555W, and F675W filers. Contours shows emission from the 0.9~mm continuum and $^{12}$CO J=3--2, $^{13}$CO J=3--2, and CS J=7--6 lines (\textit{from left to right}). HST data were retrieved from the Hubble Legacy Archive and are associated with Proposal 9315 (PI K. Noll). Contour levels are marked in each panel. Features in the scattered light/HST image are annotated in the first panel.}
\label{fig:NIR_vs_lines}
\end{figure*}

Next, we compare the HST images to the mm dust and molecular gas distribution. Since our ALMA observations were taken more than two decades after the HST observations, we first had to account for the source proper motion (see Appendix \ref{sec:app:proper_motion} for a description of this process). Once corrected, we then directly overlaid the ALMA data onto the scattered light images in Figure \ref{fig:NIR_vs_lines}.

As shown in Figure \ref{fig:NIR_vs_lines}, the millimeter continuum coincides with the dark lane in the HST image, consistent with large grains being concentrated in the disk midplane. The $^{12}$CO radial and vertical morphology closely follows that of the small dust, but with small-scale differences in their flaring structure. The $^{13}$CO emission lies at lower heights than the small dust, while CS is concentrated closer to the midplane. This stratification is consistent with observations of other disks, where the small grains typically reside at the $^{12}$CO emission surface or between the $^{12}$CO and $^{13}$CO emitting layers \citep[e.g.,][]{Rich21, Wolff21, Flores21, Law22, Law23}. Thus, in this sense, the relative gas-to-dust vertical structure in the GoHam disk appears typical.

\subsection{Origins of the Millimeter Dust Asymmetry} \label{sec:vortex_continuum}

Millimeter continuum asymmetries, including prominent crescent- or arc-shaped structures \citep[e.g.,][]{vanderMarel13, Perez14, Casassus15, Boehler18}, have been observed in a number of protoplanetary disks. However, such strong, one-sided dust morphologies as seen in GoHam appear to be relatively uncommon compared to the axisymmetric rings and gaps that are more frequently observed in ALMA surveys \citep[e.g.,][]{Andrews18, Long18}. In this context, the presence of radially-extended dust in the southern part of the GoHam disk suggests a localized, non-axisymmetric structure or perturbation capable of concentrating and redistributing millimeter-sized grains  \citep[e.g.,][]{Dullemond18}. The spatial coincidence of this southern continuum enhancement with the previously-identified PAH deficit \citep{Berne15} and $^{13}$CO excess near GoHam~b further supports a connection between the millimeter dust asymmetry and the proposed gas over-density.

The southern continuum enhancement also differs from the northern disk in its vertical morphology. It extends to larger radii, but has a narrower vertical distribution, which may reflect a genuine difference in the vertical distribution of millimeter-sized grains, rather than solely an azimuthal variation in surface density. A localized perturbation could modify the settling or redistribution of large grains through gravitational torques, spiral shocks, or changes in the local pressure structure \citep[e.g.,][]{Boley07,Ilee17}. Alternatively, the observed difference in vertical extent could result from variations in dust temperature or optical depth along the line of sight, particularly given the nearly edge-on geometry. The coincidence of the narrower southern continuum structure with the gap-like feature in scattered light suggests that the asymmetry may involve both the large- and small-grain populations.

One possible origin is a GI-driven structure associated with the proposed GoHam~b gas over-density. Massive disks near the fragmentation threshold are expected to develop spiral arms, over-densities, and, in some cases, bound fragments that generate local pressure maxima capable of efficiently trapping millimeter-sized grains \citep[e.g.,][]{Rice04, Rice06, Dipierro15}. Numerical simulations demonstrate that such GI-induced dust concentrations can be azimuthally localized and radially extended, particularly in systems with high disk-to-stellar mass ratios \citep[e.g.,][]{Gibbons12, Hall16, Forgan18}. The spatial coincidence of this southern dust enhancement with localized SO emission and an inferred gas over-density (Figure \ref{fig:goham_b_vs_dust}) is consistent with this picture, in which gas and dust asymmetries arise from the same underlying perturbation \citep[e.g.,][]{Ilee17, Hall20}. In further support of this idea, \citet{Baehr22} found that GI-driven spirals produced shocks at 1-2 scale heights above the midplane, which were typically spaced between the dense spirals. This prediction matches the local, but widespread, nature of the SO features (Figure \ref{fig:SO_detection}).  In this interpretation, GoHam~b represents either a bound fragment or a pre-fragment over-density embedded in the disk, capable of trapping large grains and producing the observed southern continuum excess. GI-driven spirals and associated density perturbations can also produce non-axisymmetric structure in the small-grain distribution and hence in scattered light \citep[e.g.,][]{Dong15, Pohl15}. However, they are not necessarily expected to produce a gap on only one side of the disk (Figure \ref{fig:goham_b_vs_dust}), leaving its origin uncertain.

An alternative explanation is an eccentric disk geometry, similar to that recently proposed for the comparably massive and extended Dracula’s Chivito disk \citep{Lovell25}. In that system, an eccentric model reproduces the observed one-sided millimeter continuum excess without invoking compact dust traps. Eccentric disks naturally generate azimuthal surface density contrasts and enhanced emission near apocenter due to orbital crowding and reduced radial drift \citep[e.g.,][]{Ataiee13, Ragusa17, Lovell23}. Such eccentricity may arise from self-gravity in massive disks or interactions with embedded companions \citep[e.g.][]{Papaloizou01, Kley08}. A similar mechanism could plausibly operate in GoHam given its large disk mass. An eccentric disk could produce asymmetric scattered-light emission through azimuthal variations in the disk geometry and surface density, although the relationship between eccentricity and scattered-light morphology is not necessarily straightforward \citep[e.g.,][]{Lovell25}.

Other mechanisms, such as vortices triggered by the Rossby Wave Instability \citep[e.g.,][]{Lyra09, Baruteau16}, planet-disk interactions \citep[e.g.,][]{Zhu14}, or misaligned inner disks that introduce shadowing and asymmetric heating \citep[e.g.,][]{Min17, Facchini18_shadow}, can also produce asymmetric dust concentrations. Of these, the misaligned inner disk scenario may provide the most natural explanation for the observed scattered-light morphology, as the apparent gap in small grains is confined to one side of the disk. Similar one-sided gaps arise in models of warped inner disks \citep{Facchini18_shadow}, but vortex-driven dust trapping generally requires an axisymmetric gas pressure structure and would not produce the observed one-sided scattered-light gap \citep[e.g.,][]{Min17, Hammer19}. 

To discriminate between these scenarios, high-resolution multi-band continuum observations that better probe the full dust size distribution, including those in the cm, are critical. In particular, resolved spectral index maps can constrain spatial variations in the dust opacity spectral index and hence the relative abundance of large grains, providing a test of whether the southern continuum enhancement is associated with localized dust trapping and concentration \citep[e.g.,][]{Birnstiel18, Tazzari21}. Such measurements across the GoHam disk, especially in the vicinity of the asymmetric southern disk, will be particularly constraining. Observations capable of probing the innermost disk geometry will also be required to assess whether shadowing contributes to the dust asymmetry.

\subsection{SO as a Tracer for Fragment-Driven Heating} \label{sec:SO_emission}

\begin{figure*}
\centering
\includegraphics[width=\linewidth]{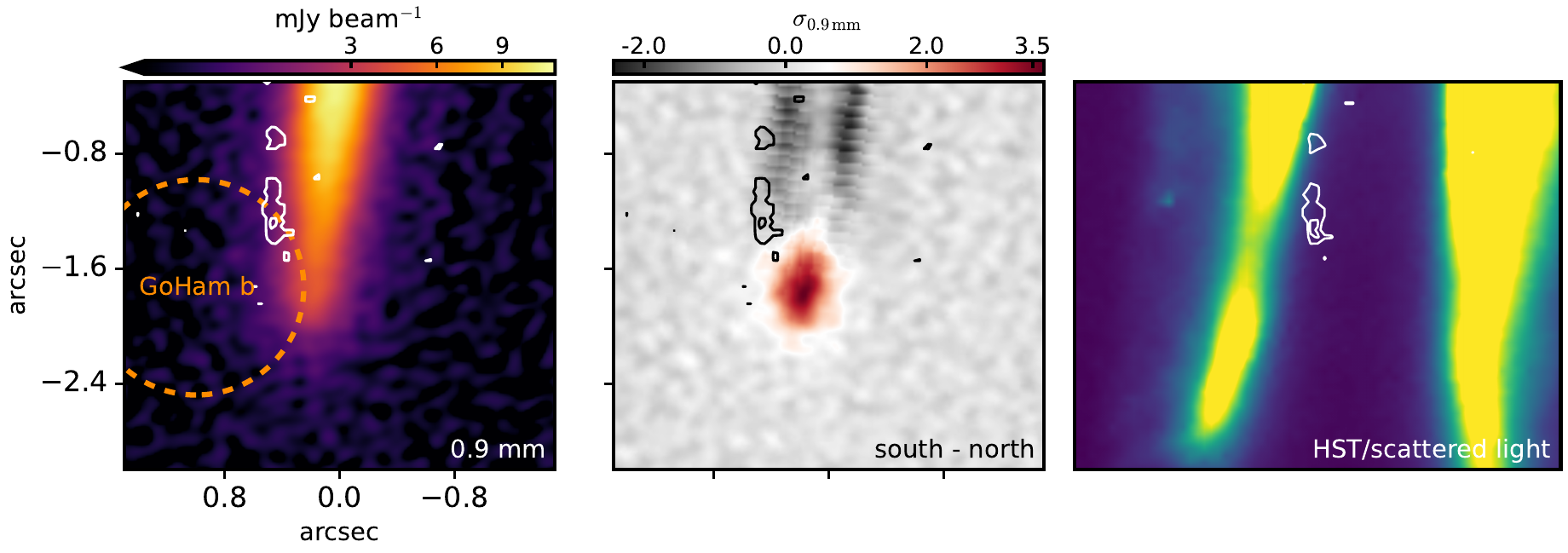}
\caption{A zoom-in on the sourthern part of the GoHam disk. The 0.9~mm continuum emission (\textit{left}) and associated residuals (in units of SNR) from reflecting and subtracting north from south halves of the disk (\textit{middle}), and HST scattered light image (\textit{right}). A color stretch has been applied to the HST image to highlight faint disk features. The SO peak intensity contours are overlaid on each panel. The approximate location of GoHam~b \citep{Bujarrabal09} is overlaid as a dotted orange circle in the 0.9~mm image.}
\label{fig:goham_b_vs_dust}
\end{figure*}

The SO emission is confined to a narrow arc along the eastern edge of the millimeter continuum and exhibits a pronounced north-south brightness asymmetry. Its brightest emission coincides with the southern dust enhancement, the previously-proposed GoHam~b gas over-density, and the localized deficit of small grains (Figure \ref{fig:goham_b_vs_dust}), suggesting that the sulfur chemistry is tracing the same localized perturbation responsible for the broader disk asymmetry.

This localized SO emission may represent a chemical signature of fragment-driven heating. In models of GI fragmentation, spiral arms and bound clumps generate localized compressional heating and shocks that can elevate gas temperatures above the sublimation thresholds of sulfur-bearing ices \citep{Boley07, Ilee11, Ilee17}. Because sulfur is efficiently locked in grain mantles at low temperatures \citep[e.g.,][]{Aikawa02}, even modest temperature enhancements can rapidly return sulfur-bearing species to the gas phase, initiating gas-phase reactions that produce SO. Laboratory measurements and chemical models indicate that sulfur-bearing species, such as SO, thermally desorb at dust temperatures of ${\sim}$20-70~K, depending on binding energies and ice composition \citep{Wakelam17, Perrero22}. At typical outer-disk densities (n$_{\rm{H}}$~${\gtrsim}$~10$^7$~cm$^{-3}$), adsorption timescales onto micron-sized grains are on the order of 10$^2$-10$^3$~yr \citep[e.g.,][]{Hollenbach09, Walsh10}, comparable to, or shorter, than orbital periods at radii of ${\sim}$100-200~au. Thus, sustained gas-phase SO emission requires either ongoing heating or a sufficiently recent thermal perturbation, making it unlikely that the observed one-sided enhancement in the GoHam disk reflects a persistent abundance asymmetry in the absence of such a perturbation. SO may also trace chemical signatures of nascent dynamical collapse. If sulfur is locked in grains in the form of refractory chains, as suggested by observations in the ISM \citep{Druard12} and comet 67P \citep{Mahjoub24}, these species could be liberated through sputtering driven by dust-dust collisions in collapsing gas. The released sulfur may then rapidly react with oxygen, itself liberated by sputtering, leading to enhanced SO production \citep{Tielens94}.

Importantly, the entirety of the SO-emitting gas does not need to be gravitationally bound to a fragment. In GI models, fragments are embedded in the differentially rotating disk flow, and only gas within the fragment’s Hill sphere remains permanently bound \citep[e.g.,][]{Boley07, Ilee11}. For a representative fragment mass of a few Jupiter masses at the location of GoHam~b, the Hill radius is ${\approx}$10-20~au, smaller than the observed azimuthal extent of the SO emission (Figure \ref{fig:SO_detection}). Gas passing through the fragment’s vicinity, or its spiral wake, can be heated and chemically processed before continuing on approximately Keplerian orbits. Given the ${\sim}$10$^2$-10$^3$~yr adsorption timescale, gas-phase SO can persist for a significant fraction of an orbit as it shears azimuthally away from the heating source. In this picture, the extended, arc-like morphology may represent a chemical wake trailing a localized perturbation rather than emission strictly co-spatial with a bound clump. The bound clump itself would instead be traced by the brightest SO emission near GoHam~b. 

GI-driven perturbations are also expected to modify the disk vertical morphology. Three-dimensional simulations show that spiral arms and fragments can induce local vertical expansion, shocks, and temperature inversions, perturbing the canonical layered disk structure \citep{Boley07, Forgan11, Ilee17}. Such vertical restructuring can lift material into warmer layers, where ice sublimation and gas-phase reactions are enhanced. As GI fragments can accrete material over the vertical extent of the disk, this may cause the vertical displacement of gas and entrained small grains, which is potentially consistent with the gap in small dust seen in the right panel of Figure \ref{fig:goham_b_vs_dust}. The SO emission may not only trace localized heating but also indicate fragment-induced vertical restructuring of the disk.

However, SO emission alone does not uniquely identify gravitational instability. Sulfur oxides are known to be enhanced in shocks and irradiated environments \citep[e.g.,][]{Pineau93, Wakelam05, Garufi22, vanGelder22, Liu25}, and localized heating that is not related to bound fragments could similarly produce SO emission. For instance, the emission arises near the outer edge of the millimeter dust disk, where reduced continuum optical depths and increased penetration of external or stellar UV photons may promote the desorption of sulfur-bearing ices and boost gas-phase SO abundances \citep{Martin_Domenech25}. Similarly, an asymmetric UV/X-ray irradiation field, potentially produced by a modest disk warp or time-variable stellar magnetospheric hot spots, could preferentially heat one side of the disk and generate a localized sulfur enhancement \citep{Young21}. Another possibility is that SO traces an infall streamer impacting the outer disk \citep[e.g.,][]{Garufi22, Huang23, Speedie25}. Such an interaction could produce localized shocks and heating capable of enhancing SO emission, while simultaneously perturbing the dust distribution. In this context, GoHam shares some similarities with the AB~Aur disk, where a localized SO enhancement and millimeter dust asymmetry have been linked to late infall \citep[e.g.,][]{Tang12, Dutrey24, Speedie25}. However, AB~Aur exhibits a prominent scattered-light streamer associated with the inferred impact location, but no comparable structure is evident in GoHam (see Section \ref{sec:scattered_light_morph}; Figure \ref{fig:NIR_vs_lines}). In general, it is not clear whether any of these alternative scenarios can explain the spatial coincidence of the SO emission, GoHam~b over-density, millimeter continuum asymmetry, and gap in the small-grain dust distribution.

If the SO enhancement does trace fragment-driven heating, this interpretation yields testable predictions. Fragmenting disk models predict that localized heating will enhance other sulfur-bearing species, including OCS, SO$_2$, and H$_2$S, with peak abundances concentrated near the fragment rather than smoothly distributed along spiral arms \citep{Ilee17}. Molecular tracers of hot gas and local ice sublimation, such as H$_2$O, H$_2$CO, and even larger organic species like CH$_3$OH, should likewise show co-spatial excitation enhancements \citep[e.g.,][]{Walsh14}. A spatially-resolved, multi-line excitation analysis of the local gas conditions around GoHam~b will thus be critical for distinguishing fragment-driven thermochemistry from other dynamical processes.

\section{Conclusions} \label{sec:conlcusions}

We present high-angular-resolution ALMA observations of the edge-on GoHam disk that reveal highly asymmetric SO emission, likely associated with the earliest phases of wide-separation, giant planet formation or disk fragmentation. The edge-on geometry allows us to place this chemical signature in the broader context of the gas, dust, and dynamical disk structure. We conclude the following:

\begin{enumerate}
    \item We detect SO emission as a narrow arc on the eastern side of the disk with a pronounced north-south brightness asymmetry, peaking near the previously inferred GoHam~b gas over-density. This morphology is consistent with localized heating associated with an embedded fragment or forming giant planet, which is expected to enhance gas-phase SO abundances.
    \item We identify an asymmetry in the millimeter continuum along the disk major axis, characterized by a larger radial extent, but narrower vertical distribution, in the southern part of GoHam's disk. The difference in vertical extent may indicate an asymmetric distribution of millimeter-sized grains, although variations in optical depth and temperature could also contribute to the observed morphology. This localized non-axisymmetric structure is potentially driven by GI or an eccentric disk geometry. The southern continuum enhancement is also spatially coincident with a gap-like feature in the scattered-light surface near GoHam~b.
    \item We map the vertical emission structure of $^{12}$CO ($z/r\sim0.5$-0.6), $^{13}$CO ($z/r\sim0.4$), and CS ($z/r\sim0.15$). Each species arises progressively closer to, but remains above, the disk midplane as traced by the mm dust continuum. The $^{12}$CO emission surface is lies at approximately the same altitude as the scattered-light surface. Each molecule shows some degree of spatial asymmetry along the major, and in some cases, minor disk axis, suggesting departures from an axisymmetric disk structure that may be associated with dynamical perturbations or disk winds.
    \item We identify extended, non-Keplerian $^{12}$CO and $^{13}$CO emission at large radii toward the north half of the GoHam disk. The morphology and kinematics of these features are broadly consistent with a disk wind, although IFU spectro-imaging observations (i.e., with JWST NIRSpec, MIRI) and dedicated modeling will be required to constrain its physical origin and properties.
    \item Using a dust-extinction-map-based technique, we derive a revised distance of 139~$\pm$~24~pc, placing GoHam in the outskirts of the Scorpius-Centaurus association. We also infer a dynamical stellar mass of 2.2 $\pm$ 0.5 M$_{\odot}$ by fitting the kinematics of $^{12}$CO, $^{13}$CO, and CS.
\end{enumerate}

\begin{acknowledgments}
The authors thank the anonymous referee for valuable comments that improved the content and presentation of this work. This paper makes use of the following ALMA data: ADS/JAO.ALMA\#2022.1.00269.S. ALMA is a partnership of ESO (representing its member states), NSF (USA) and NINS (Japan), together with NRC (Canada), MOST and ASIAA (Taiwan), and KASI (Republic of Korea), in cooperation with the Republic of Chile. The Joint ALMA Observatory is operated by ESO, AUI/NRAO and NAOJ. The National Radio Astronomy Observatory is a facility of the National Science Foundation operated under cooperative agreement by Associated Universities, Inc. The Submillimeter Array is a joint project between the Smithsonian Astrophysical Observatory and the Academia Sinica Institute of Astronomy and Astrophysics and is funded by the Smithsonian Institution and the Academia Sinica.

Support for C.J.L. was provided by NASA through the NASA Hubble Fellowship grant No. HST-HF2-51535.001-A awarded by the Space Telescope Science Institute, which is operated by the Association of Universities for Research in Astronomy, Inc., for NASA, under contract NAS5-26555. K.M. was supported by HST-GO-17751, JWST-GO-1905 and JWST-GO-3523. T.J.H. acknowledges a Dorothy Hodgkin Fellowship and UKRI guaranteed funding for a Horizon Europe ERC consolidator grant (EP/Y024710/1). D.~S. was funded by the Deutsche Forschungsgemeinschaft (DFG, German
Research Foundation) – project number: 550639632. We would also like to thank Joshua Lovell for useful discussions that improved this manuscript.
\end{acknowledgments}

%

\facilities{ALMA, HST (WFPC2), SMA}


\software{Astropy \citep{astropy_2013,astropy_2018}, \texttt{bettermoments} \citep{Teague18_bettermoments}, CASA \citep{McMullin_etal_2007, CASATeam20}, \texttt{cmasher} \citep{vanderVelden20}, \texttt{emcee} \citep{Foreman13}, \texttt{GoFish} \citep{Teague19JOSS}, \texttt{keplerian\_mask} \citep{Teague20_Keplerian_mask}, Matplotlib \citep{Hunter07}, NumPy \citep{vanderWalt_etal_2011}, RADMC-3D \citep{Dullemond12}, \texttt{scipy} \citep{Virtanen_etal_2020}}



\clearpage

\appendix

\section{Revised Distance to GoHam} \label{sec:app:revised_distance}

\subsection{GoHam as a Member of the Nearby Sco-Cen Association}

In the solar neighborhood, distances to stars can often be determined accurately from trigonometric parallaxes. However, due to its large apparent size and nearly edge-on geometry, no reliable \textit{Gaia} parallax distance has been determined for GoHam. Instead, we estimate its distance by leveraging three-dimensional maps of interstellar dust extinction, which can be used as indirect distance indicators for nearby (${\lesssim}1$~kpc) stellar sources. When combined with information on the spatial distribution of young stellar populations, this approach allows us to constrain the most likely distance to GoHam.

We analyze the three-dimensional dust extinction maps presented by \citet{Edenhofer+2024} to identify interstellar structures along the line of sight toward GoHam and to determine the distance to the nearest dense dust cloud that could plausibly host its formation site. In a second step, we compare these results to the spatial distribution of nearby young stellar objects (YSOs) with well-determined \textit{Gaia} distances, using the recent SPYGLASS compilation by \citet{Kerr+2023}, to assess whether GoHam is associated with a known nearby star-forming population.

Distance determinations based on interstellar extinction have previously been applied successfully to infrared dark clouds \citep[e.g.,][]{Marshall+2009} and star-forming regions \citep[e.g.,][]{Kenyon+1994, Cao+2023}, and have been found to agree with maser parallax distances at the $\sim$85-100\% level outside of the crowded Galactic Center \citep{Foster+2012}. This method implicitly assumes that the target source is young enough to remain spatially associated with its natal environment. For GoHam, this assumption is well justified, as it exhibits clear signatures of youth, including a circumstellar disk in Keplerian rotation, and a stellar luminosity consistent with that of a pre-main-sequence (PMS) A-type star \citep{Bujarrabal08}.

\begin{figure}[h!]
    \centering
    \includegraphics[width=0.7\textwidth]{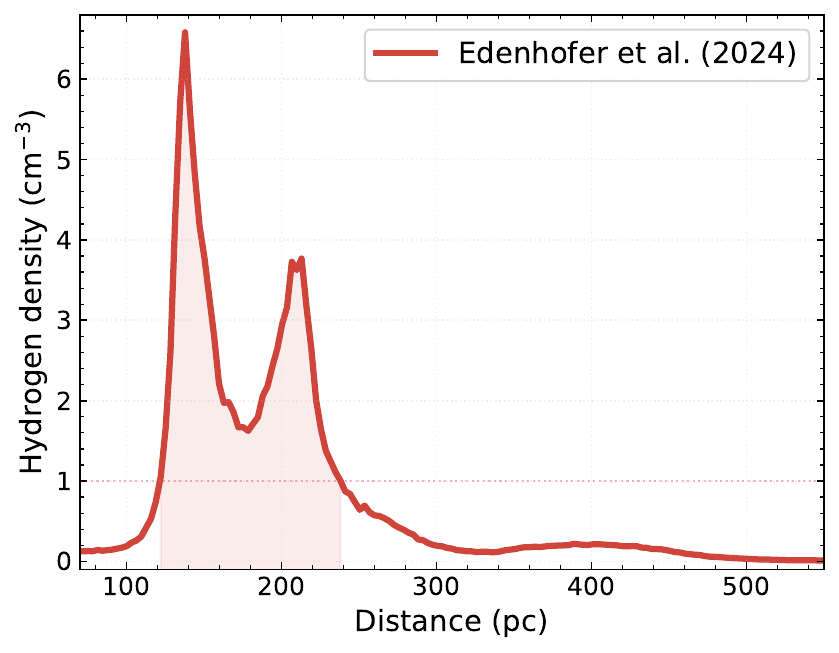}
    \caption{Reconstructed three-dimensional distribution of the interstellar hydrogen volume density along the line of sight toward GoHam, derived from the dust extinction maps of \citet{Edenhofer+2024}. The red-shaded regions highlight distances at which the hydrogen density exceeds $n_\mathrm{H}=1\,\mathrm{cm}^{-3}$, which \citet{Zucker+2021} adopt as the minimum threshold for a molecular cloud.}
    \label{fig:Gaia_dist}
\end{figure}

We queried the \citet{Edenhofer+2024} dust extinction maps around GoHam's galactic coordinates ($l=359.7222^\circ$, $b=-6.0207^\circ$), using a search radius of $2^\circ$. Following the methodology of \citet[][see their Section~2.1.1]{Zucker+2021}, we converted the differential extinction per parsec provided by the maps into a hydrogen volume density. Adopting the relation between extinction and column density from \citet{Draine2009}, i.e., $\frac{A_\mathrm{G}}{N(H)} = 4 \times 10^{-22}\,\mathrm{mag\,cm^2}$, we obtain $n_{\rm{H}} \approx 1653\,\mathrm{cm}^{-3} \times s_\mathrm{E}$, where $s_\mathrm{E}$ denotes the differential extinction per parsec from the \citet{Edenhofer+2024} maps. Consistent with \citet{Zucker+2021}, we classify regions with $n_\mathrm{H} \geq 1\,\mathrm{cm}^{-3}$ as molecular clouds.

Figure~\ref{fig:Gaia_dist} shows the resulting mean hydrogen density profile along the line of sight toward GoHam. This analysis reveals a broad candidate distance range between $\sim$122~pc and 238~pc, with two more prominent density enhancements located at approximately 128-166~pc and 188-222~pc. Based on the extinction data alone, however, it is not possible to uniquely associate GoHam with either structure, as both correspond to dust clouds massive enough to plausibly host star formation.

To further constrain the distance, we examined the spatial distribution of nearby young stars using the SPYGLASS catalog of \citet{Kerr+2023}, which provides an expanded census of YSOs younger than 50~Myr with accurate \textit{Gaia} distances within 1~kpc of the Sun. We searched for young stellar populations in cones centered on GoHam's sky position with radii of $2^\circ$, $1^\circ$, $0.5^\circ$, $0.2^\circ$, and $0.05^\circ$, under the assumption that YSOs formed in the same molecular cloud as GoHam should be spatially clustered on the sky. For reference, the Orion Nebula Cluster spans roughly $1^\circ \times 1^\circ$ on the sky.

\begin{figure}[h!]
    \centering
    \includegraphics[width=0.7\textwidth]{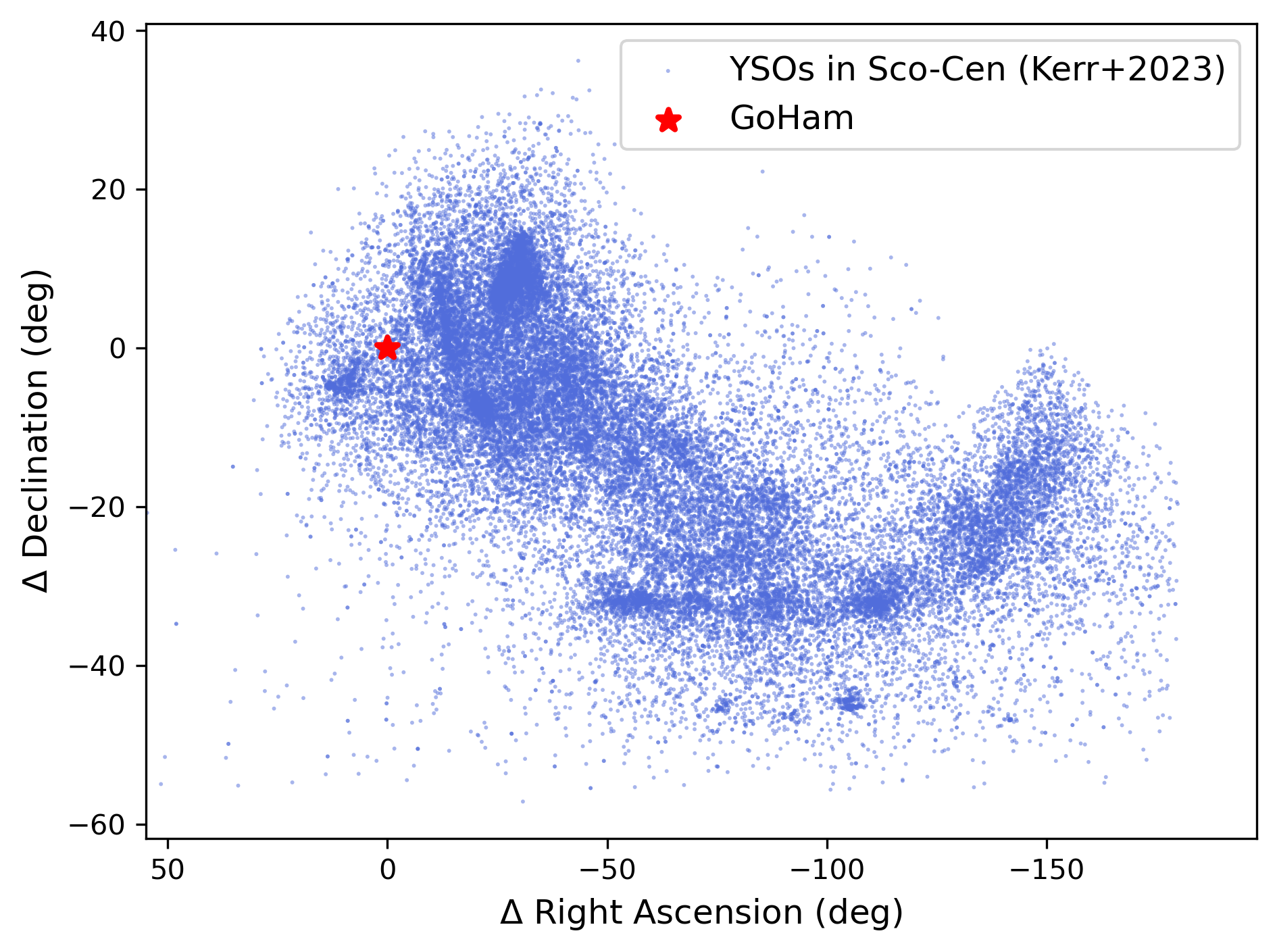}
    \caption{The positions of known young stellar objects (blue dots) are shown together with the location of GoHam (red star), illustrating its position toward the outskirts of the association.}
    \label{fig:goham_pos}
\end{figure}

We find that all PMS stars within the $2^\circ$ search cone are associated with the Scorpius–Centaurus (Sco-Cen) association. \citet{Kerr+2023} report a mean distance of $139\pm24\,\mathrm{pc}$ and an age of $\sim$17~Myr for this population. This distance coincides well with the nearer peak in the hydrogen density distribution identified in Figure~\ref{fig:Gaia_dist}. Inspecting GoHam's position relative to the other PMS members of Sco-Cen, we find that it lies on the outskirts of the association (Figure \ref{fig:goham_pos}). We therefore adopt a distance of $\approx$139~pc as the most likely distance to GoHam.

\subsection{Implications for Potential GI in GoHam}

This revised distance of 139~pc (from~250 pc) reduces the inferred disk mass by a factor of ${\approx}$0.3 and correspondingly increases the Toomre~Q estimate, since the outer disk is now physically smaller and warmer, leading to a net increase in Q by a factor of a few relative to previous estimates \citep{Berne15}. While this revision weakens the case for global gravitational instability, it does not exclude the possibility that self-gravity may still operate locally at large radii, where the surface density remains comparatively high.

Independent, order-of-magnitude constraints based on the high-critical density \citep[${\sim}$10$^6$-10$^7$~cm$^{-3}$;][]{Shirley15} CS J=7--6 line suggest outer disk surface densities on the order of ${\sim}$0.2-0.3~g~cm$^{-2}$ and a total gas mass of ${\approx}$0.06-0.07~M$_{\odot}$ (for plausible radial distributions of the gas surface density). These estimates remain close to the regime where GI could still be potentially relevant if the outer disk is sufficient cold. Given the strong dependence on temperature structure, geometry, and excitation assumptions, we defer a definitive reassessment of the disk mass and stability to future detailed radiative transfer modeling.

\section{Continuum Imaging and Fitting} \label{sec:app:robust_continuum_imaging}

\begin{figure*}[b]
\centering
\includegraphics[width=\linewidth]{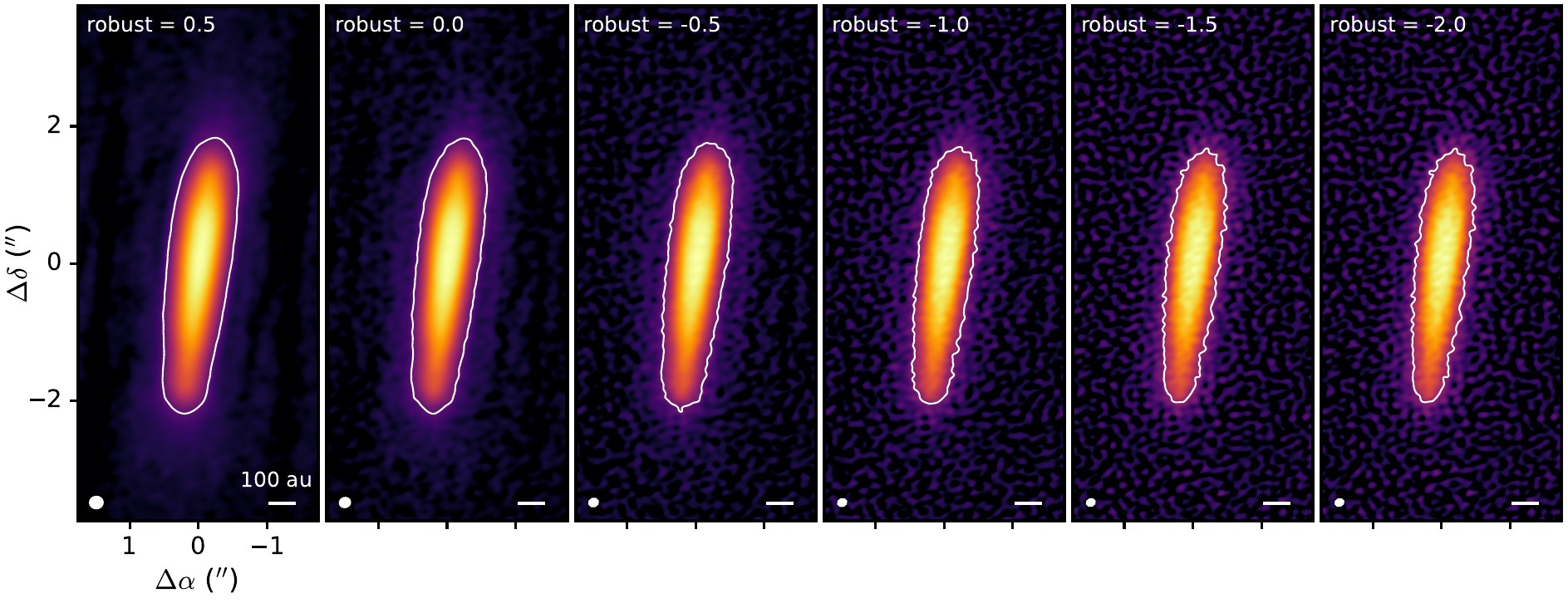}
\caption{Gallery of 0.9~mm continuum images of GoHam produced with varying \texttt{robust} values. White contours indicate the 3$\times$RMS level in each image.}
\label{fig:continuum_robust_gallery}
\end{figure*}

Figure \ref{fig:continuum_robust_gallery} shows 0.9~mm continuum images of the GoHam disk produced with different $\texttt{robust}$ values. The north-south mm dust asymmetry persists in all imaging choices. We adopted the \texttt{robust}=$-$0.5 image as our fiducial image, as this provides the ideal compromise between maximizing angular resolution and not resolving out the major axis, as in the images generated with lower \texttt{robust} values.

To further constrain the mm dust structure, we employed radiative transfer modeling. We adopted the model presented in \citet{Lin23}, where the full definition of the fitted parameters and functional forms can be found. We used \texttt{emcee} \citep{Foreman13} to conduct the parameter search by minimizing the difference between the matched model and observations. We used 3000 walkers and 500 burn-in steps to obtain the final best-fit parameters. We took the median of the posterior and 16th-84th percentiles as the best-fit and associated uncertainties, respectively. 

\begin{figure}[!t]
\centering
\includegraphics[width=0.75\linewidth]{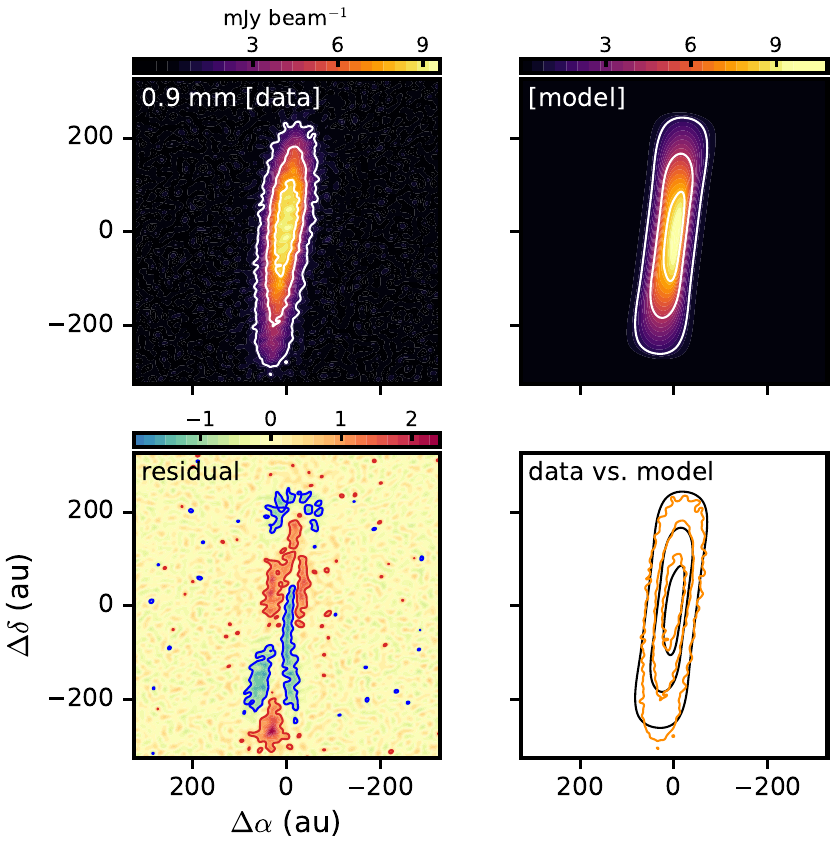}
\caption{Continuum radiative transfer fitting for the GoHam disk. The 0.9~mm continuum image, best-fit model, residuals, and contour overlays \textit{(left-to-right, top-to-bottom)}. The residuals are shown in units of $\sigma$.}
\label{fig:continuum_fit_emcee}
\end{figure}

Figure \ref{fig:continuum_fit_emcee} shows the best-fit and residuals. While this fitting produces a better fit than a simple 2D model and allows for improved geometric constraints, it does not yield a residual-free fit, confirming that the assumption of a symmetric disk is inappropriate for the complex mm dust structure present in the GoHam system. Future, non-parametric model fitting is needed to properly capture the full complexity of this disk.

\clearpage

\section{Gallery of Rotation and Peak Intensity Maps} \label{sec:app:peak_rotation}

\begin{figure}[!h]
\centering
\includegraphics[width=0.70\linewidth]{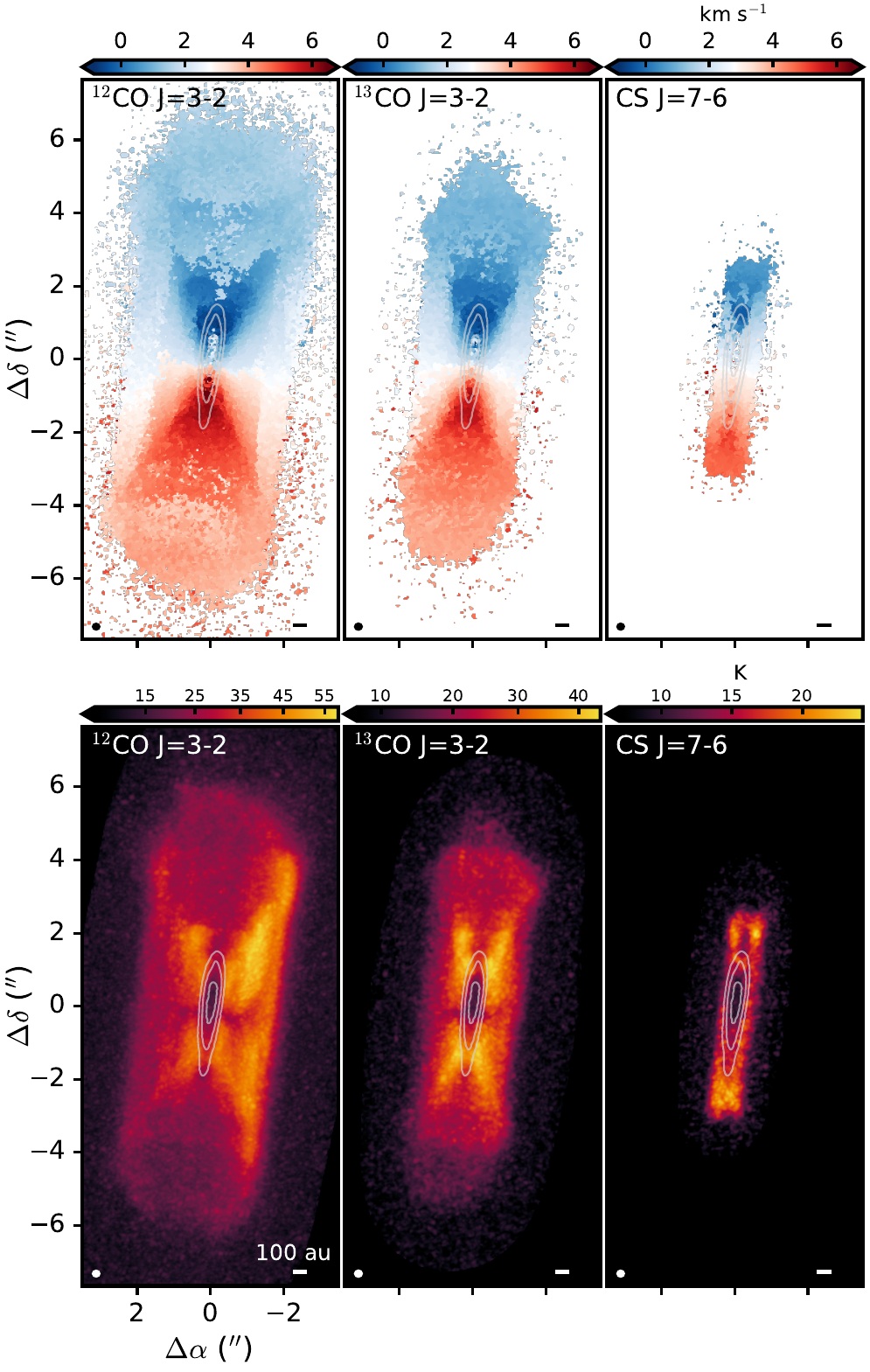}
\vspace{-6pt}
\caption{Rotation \textit{(top row)} and peak intensity \textit{(bottom row)} maps of $^{12}$CO J=2--1, $^{13}$CO J=2--1, and CS J=7--6 in the GoHam disk. The 0.9~mm continuum is overlaid in contours. The synthesized beam and a scale bar indicating
100~au is shown in the lower left and right corner, respectively, of each panel}
\label{fig:Fnu_v0_gallery}
\end{figure}

Figure \ref{fig:Fnu_v0_gallery} shows a gallery of rotation and peak intensity maps of $^{12}$CO, $^{13}$CO, and CS lines in the GoHam disk.

\section{Approximate Proper Motion Adjustments} \label{sec:app:proper_motion}

Given the two-decade time baseline between the HST observations (2002) and new ALMA data (2022-2024), we need to assess whether GoHam exhibits measurable proper motion before directly comparing the datasets. This determination is complicated by the edge-on geometry of the system, which precludes reliable astrometric measurements from \textit{Gaia}. Furthermore, the ALMA images lack background stars that can be registered against the HST field, and no intermediate-epoch HST imaging is available.

To estimate the system’s motion, we instead used archival SMA observations obtained in 2006 \citep{Bujarrabal08}. We first analyzed the 1.3 mm continuum data using the same \texttt{imfit} procedure described in Section~\ref{sec:continuum_results}. The disk major axis was spatially-resolved in the SMA data, allowing for a measurement of a relative Decl. shift of ${\approx}$30~mas~yr$^{-1}$ between the SMA and ALMA epochs. Due to the modest angular resolution of the SMA observations (${\sim}$1$^{\prime \prime}$), the disk minor axis was not resolved in continuum emission. We therefore used the more spatially-extended SMA $^{12}$CO J=2--1 emission to estimate the R.A. motion by comparing the vertical edges of iso-velocity contours with those observed in the ALMA $^{12}$CO J=3--2 data, yielding an offset of ${\approx}$14 mas~yr$^{-1}$. Although these measurements use different $^{12}$CO transitions, both lines are expected to trace comparable emitting heights in disks \citep{Law23}.

These measurements imply a total proper motion of 33.1 mas~yr$^{-1}$, consistent with Sco-Cen kinematics \citep[${\approx}$25–40 mas~yr$^{-1}$;][]{Wright18, Luhman20}. We applied these offsets when aligning the multi-epoch datasets in Section~\ref{sec:height_compar}, but note that the derived motion should be regarded as approximate pending future high-precision astrometry.

\clearpage

\bibliography{GoHam}{}
\bibliographystyle{aasjournalv7}



\end{document}